%% file: emu_paper.tex
\documentclass[fleqn,usenatbib]{mnras}
\RequirePackage{snapshot}

\usepackage{newtxtext,newtxmath}
\usepackage{booktabs}	
\usepackage{multirow}	

\usepackage[utf8]{inputenc}
\usepackage[T1]{fontenc}
\usepackage{ae,aecompl}

\usepackage{graphicx}	
\usepackage{amsmath}	
\usepackage{amssymb}	
\usepackage{bm}
\usepackage[dvipsnames]{xcolor}
\usepackage{grffile}

\newcommand{\Msunh}{\ensuremath{ M_{\odot} h^{-1}}}
\newcommand{\hMpc}{\ensuremath{h^{-1} \, \text{Mpc}}}
\newcommand{\xigg}{\ensuremath{ \xi_{\text{gg}} }}
\newcommand{\xigm}{\ensuremath{ \xi_{\text{gm}} }}
\newcommand{\ximm}{\ensuremath{ \xi_{\text{mm}} }}
\newcommand{\wgm}{\ensuremath{ w_{\text{gm}} }}
\newcommand{\Pgg}{\ensuremath{ P_{\text{gg}} }}
\newcommand{\Pgm}{\ensuremath{ P_{\text{gm}} }}
\newcommand{\Pmm}{\ensuremath{ P_{\text{mm}} }}
\newcommand{\lcdm}{$\Lambda$CDM}

\input{header.tex}

\title[Small-scale cosmology emulation]{Cosmology with galaxy-galaxy lensing on non-perturbative scales: Emulation method and application to BOSS LOWZ}

\author[B. D. Wibking et al.]{Benjamin D. Wibking$^{1}$\thanks{Current institution: Research School of Astronomy and Astrophysics, Australian National University. E-mail: benjamin.wibking@anu.edu.au},
David H. Weinberg$^{1}$,
Andrés N. Salcedo$^{1}$,
Hao-Yi Wu$^{1}$,
\newauthor
Sukhdeep Singh$^{2}$,
Sergio Rodríguez-Torres$^{3,\,4}$,
Lehman H. Garrison$^{5}$,
\newauthor
and Daniel J. Eisenstein$^{5}$ \\
$^{1}$Dept. of Astronomy and Center for Cosmology and AstroParticle Physics, Ohio State University, 140 W 18th Ave, Columbus, OH, USA\\
$^{2}$Berkeley Center for Cosmological Physics, University of California, Berkeley, Berkeley, CA, USA\\
$^{3}$Instituto de Astrofísica de Canarias, s/n, E-38205, La Laguna, Tenerife, Spain\\
$^{4}$Departamento de Física Teórica M8, Universidad Autónoma de Madrid (UAM), Cantoblanco, E-28049, Madrid, Spain\\
$^{5}$Harvard-Smithsonian Center for Astrophysics, 60 Garden St., MS-10, Cambridge, MA 02138\\
}

\date{Accepted 2019 December 3. Received 2019 December 3; in original form 2019 July 14.}

\pubyear{2019}

\begin{document}
\label{firstpage}
\pagerange{\pageref{firstpage}--\pageref{lastpage}}
\maketitle

\begin{abstract}
We describe our nonlinear emulation (i.e., interpolation) framework that combines the halo
occupation distribution (HOD) galaxy bias model with $N$-body
simulations of nonlinear structure formation, designed to accurately predict the projected clustering and galaxy-galaxy lensing signals from luminous red galaxies (LRGs) in the redshift range $0.16 < z < 0.36$ on comoving scales $0.6 < r_p < 30$ \hMpc. The interpolation accuracy is $\lesssim 1-2$ per cent across the entire physically plausible range of parameters for all scales considered.
We correctly recover the true value of the cosmological parameter $S_8 = ({\sigma_8}/{0.8228}) ({\Omega_{\text{m}}}/{0.3107})^{0.6}$ from mock measurements produced via subhalo abundance matching (SHAM)-based lightcones designed to approximately match the properties of the SDSS LOWZ galaxy sample.
Applying our model to Baryon Oscillation Spectroscopic Survey (BOSS) Data Release 14 (DR14) LOWZ galaxy clustering and galaxy-shear cross-correlation measurements made with Sloan Digital Sky Survey (SDSS) Data Release 8 (DR8) imaging, we perform a prototype cosmological analysis marginalizing over $w$CDM cosmological parameters and galaxy HOD parameters. We obtain a 4.4 per cent measurement of $S_8 = 0.847 \pm 0.037$, in $3.5\sigma$ tension with the \emph{Planck} cosmological results of $1.00 \pm 0.02$. We discuss the possibility of underestimated systematic uncertainties or astrophysical effects that could explain this discrepancy.
\end{abstract}

\begin{keywords}
cosmology -- weak lensing -- large scale structure
\end{keywords}



\section{Introduction}
Weak gravitational lensing has emerged as the most powerful probe of matter clustering in the low redshift universe, critical to testing whether cosmic acceleration is caused by a cosmological constant, by an alternative form of dark energy, or by a breakdown of General Relativity on cosmological scales. Cosmic shear measures the clustering of foreground dark matter from the correlations induced in the shape of background source galaxies. Galaxy-galaxy lensing (GGL) uses a background shear map to measure the clustering of matter around a foreground galaxy population, which can be combined with the foreground galaxy clustering itself to infer the underlying dark matter clustering. In applications to state-of-the-art weak lensing data sets, the two approaches have comparable statistical uncertainties, with systematics that are partly in common and partly distinct (e.g. \citealt{Hildebrandt_2017,DESY1KP}). Building on our previous work \citep{Wibking_2019}, this paper presents a numerical approach to modeling galaxy clustering and GGL into the deeply non-linear regime and applies it to weak lensing and galaxy clustering measurements \citep{Singh_2018} from the LOWZ galaxy sample of the Baryon Oscillation Spectroscopic Survey (BOSS; \citealt{Eisenstein_2011,Dawson_2013}), including tests on the LOWZ mock catalogs used by \cite{Singh_2018}.

On asymptotically large scales, where linear theory and scale-independent galaxy bias should be exact, one can think of galaxy clustering + GGL as measuring $\xigg = b_g^2 \ximm$ and $\xigm = b_g \ximm$, allowing cancellation of the unknown $b_g$ and inference of $\ximm$. The accuracy of this method can be improved by using higher order perturbative models of galaxy bias (reviewed by \citealt{Desjacques_2018}). However, these models break down on comoving scales below $\sim 10 \, \hMpc$, so non-perturbative models are needed to exploit clustering and GGL data on the $\sim$ Mpc scales where they are most precise.  Demands on the accuracy of model predictions will become more stringent with the completion of current generation weak lensing surveys such as the Kilo-Degree Survey (KiDS; \citealt{Hildebrandt_2017}), the Dark Energy Survey (DES; \citealt{DESY1KP}), and the Hyper-Suprime Camera Strategic Survey Program (HSC SSP; \citealt{HSC2017}), and with the advent of future surveys from the Euclid mission \citep{Laureijs_2011}, the Large Synoptic Survey Telescope (LSST; \citealt{LSST2018}), and the Wide Field Infrared Survey Telescope (WFIRST; \citealt{Dore_2019}). The dilemma of scales is already illustrated by existing analyses. Most studies of clustering + GGL on large scales infer an amplitude of matter clustering that is lower than predicted by a $\Lambda$CDM cosmological model (cold dark matter with a cosmological constant) normalized to Planck cosmic microwave background data (e.g. \citealt{Mandelbaum_2013,Hildebrandt_2017,DESY1KP}; but see \citealt{More_2015} for a compatible result). However, the significance of the discrepancy with any one data set is limited because the statistical errors on these scales are large. \cite{Leauthaud_2017} find a much stronger discrepancy on Mpc scales between measured GGL for BOSS CMASS galaxies and predictions of Planck-normalized mock galaxy catalogs tuned to CMASS galaxy clustering, but they are hesitant to draw strong conclusions because of theoretical uncertainties in the clustering models.  Recently, \cite{Singh_2018} used information down to $1 \, \hMpc$ scales with a nonparametric model of scale-dependent galaxy bias and inferred a lower amplitude of matter clustering than Planck at $> 3 \sigma$ significance. They likewise caution that uncertainties about modeling the galaxy population prohibit strong conclusions about cosmological physics.

In this work, we adopt the halo occupation distribution (HOD) model of galaxy bias \citep{Jing_1998,Peacock_2000,Seljak_2000,Scoccimarro_2001,Berlind_2002}, which is commonly used as a model of galaxy clustering on $\sim$Mpc to sub-Mpc scales (e.g., \citealt{Zehavi_2005,Zehavi_2011,Coupon_2012,Sinha_2018}) and also in combination with GGL (e.g. \citealt{Zu_2015}).  Several previous papers have advanced the idea of modeling non-linear galaxy clustering and GGL with HODs, in effect allowing the HOD to provide nuisance parameters that one can marginalize over when deriving cosmological constraints (e.g. \citealt{Yoo_2006,Cacciato_2009,Cacciato_2013,Yoo_2012}). \cite{Cacciato_2013} apply this approach to Sloan Digital Sky Survey (SDSS) data and find $\sigma_8$ and $\Omega_m$ values in good agreement with the WMAP7 results \citep{Komatsu_2011} but low compared to recent values from Planck \citep{Planck_2016}. \cite{More_2015} likewise apply this approach to CMASS galaxy clustering and Canada-France-Hawaii Telescope Lensing Survey (CFHTLS) shear catalogs and find $\sigma_8$ and $\Omega_m$ values in good agreement with both final WMAP and early Planck results. These papers have relied on analytic formulations of the HOD/halo model, which are accurate at the $\sim 5$ per cent level relative to numerical predictions from cosmological $N$-body simulations. It seems unlikely that a first-principles analytic approach can achieve the per cent-level accuracy demanded by current data sets, in part because of uncertainties in the effects of halo exclusion and scale-dependent halo bias (see e.g., \citealt{vdBosch_2013}).

By directly interpolating results from populated $N$-body simulations, on the other hand, we can compute the predictions of the halo model for the projected galaxy clustering $w_p$ and galaxy-galaxy lensing $\Delta\Sigma$ with sub-percent accuracy into the deeply non-linear regime. The main disadvantage of this approach is that it requires a large library of $N$-body simulations to sample the cosmological parameter space and many repopulations and pair counting computations of each of these simulations to sample the HOD parameter space. The range of galaxy samples that one can model is limited by the minimum mass of a well-resolved halo for a simulation of a given resolution ($\sim 300-500$ particles per halo; e.g. \citealt{Trenti_2010,Reed_2013}). In this paper, we use the AbacusCosmos suite of simulations \citep{Garrison_2017} sampling the parameters of $w$CDM cosmology, extending our previous work \citep{Wibking_2019} that used a grid of $(\sigma_8, \Omega_m)$ values within $\Lambda$CDM. A similar effort, using different simulations and a somewhat different approach to computing their predictions, has been undertaken by \cite{Zhai_2019}.

Interpolation across the outputs of simulations has become popularly known within the cosmology community as \emph{emulation} \citep{Heitmann_2009,Kwan_2015}. As with many previous efforts, we use Gaussian processes, which are commonly adopted for relatively small multidimensional datasets that are expensive to obtain and where it is desired to propagate quantitative uncertainty estimates from the input training data to the resulting predictions.\footnote{Spline interpolation, for instance, becomes increasingly difficult in high dimensions, except when implemented as a Gaussian process kernel, in which case it generally is not a good choice of kernel function.}  The underlying idea of using Gaussian process regression methods to smoothly predict the output of computer simulations is at least three decades old \citep{Sacks_1989}, and the general theory of using Gaussian processes for regression is even older \citep{Ohagan_1978}, with its original one-dimensional time-series formulation due to \cite{Kolmogorov_1941} and \cite{Wiener_1949}. To obtain sufficient accuracy for our interpolation of projected galaxy clustering $w_p$ across our parameter space, we emulate the ratio of the numerically computed $w_p$ to an approximate analytic calculation using the same cosmological and HOD parameters. For $\Delta\Sigma$, we find that direct emulation is sufficiently accurate.

Before applying our emulator to the BOSS LOWZ data, we test its ability to recover the correct cosmological parameters from the light-cone mock catalogs created by \cite{Nuza_2013} and \cite{RodriguezTorres_2016}, as used in \cite{Singh_2018}. These mock catalogs use subhalo abundance matching (SHAM) with parameters designed to reproduce the clustering and number density properties of the LOWZ samples from $z = 0.16-0.36$. Crucially, they are not created with an HOD prescription, so these tests provide at least some evidence that our parameterization is flexible enough to represent a range of possible scenarios for galaxy formation physics. We compute samples from the posterior conditioned on these mock data and show that we can recover the parameter $S_8 \propto \sigma_8 \Omega_m^{0.6}$ with bias $\leq 0.5 \sigma$, even when using scales down to $\sim 0.6 \, \hMpc$, which is the minimum usable scale because of fiber collision systematics in the clustering measurements.

In section \ref{section:emulator}, we describe the simulations and methodology used to construct our emulator and the technical details necessary to obtain our target accuracy.  Section \ref{section:covariance} describes the construction of the covariance matrix for the BOSS LOWZ analysis. Section \ref{section:mocks} presents the mock catalogs tests and section \ref{section:data} the results from application to the \cite{Singh_2018} GGL and clustering measurements. We conclude in section \ref{section:conclusions} with a discussion of the limitations of our method, proposed extensions and modifications for future work, and prospects for application to DES, HSC, WFIRST, and LSST.

We use comoving densities and distances throughout and assume flat $\Lambda$CDM for computations of these quantities, unless specified otherwise.

\section{Emulator Construction}
\label{section:emulator}

\begin{figure*}
  \includegraphics[width=\textwidth]{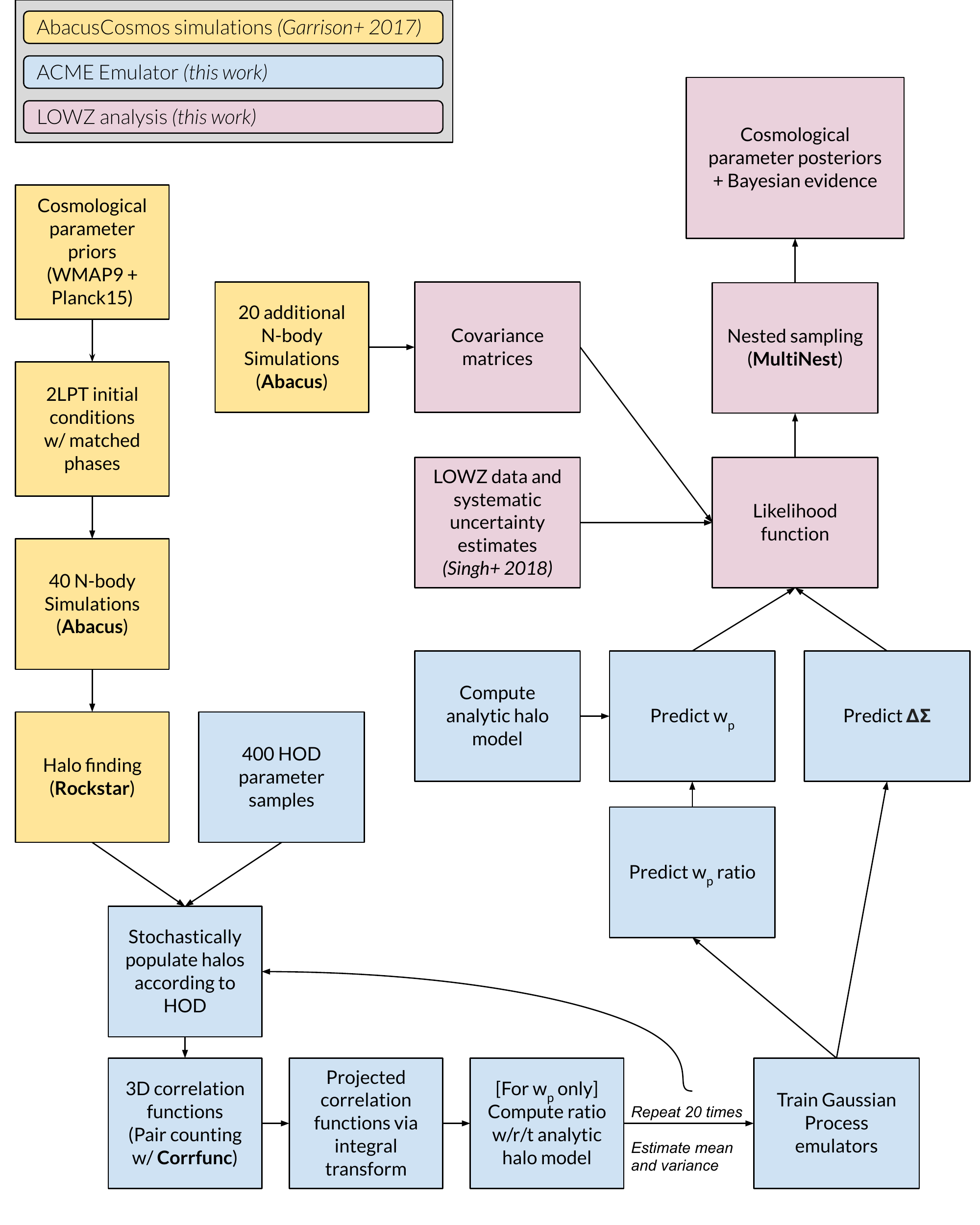}
 \caption{A high-level overview of the flow of information, from cosmological parameter priors to emulator to cosmological parameter posteriors.}
  \label{fig:flowchart}
\end{figure*}

Figure \ref{fig:flowchart} presents an overview of our emulation and inference framework. The critical inputs are the AbacusCosmos simulations, which we combine with HOD populations to train the Gaussian process emulator. We use 20 \textsc{Abacus} simulations of the fiducial cosmology given in Table \ref{table:parameters} to compute part of the covariance matrix for the BOSS LOWZ data and also to correct the mean model predictions for cosmic variance of a single simulation volume. Applying the emulator to the data with these covariances yields our cosmological parameter constraints marginalized over HOD parameters and nuisance parameters describing possible weak lensing systematics.
 
\subsection{Parameter space}

We initially chose a halo occupation distribution (HOD) parameter space designed to encompass the posterior parameter ranges for the LOWZ sample found by \cite{Parejko_2013} when fitting a simulation-based HOD model to the projected galaxy correlation function $w_p$ at fixed cosmology ($\log M_{\text{min}}=13.25 \pm 0.26$, $\sigma_{\log M}=0.43 \pm 0.25$, $\log M_0=13.27^{+0.49}_{-0.76}$, $\log M_1=14.18 \pm 0.39$, $\alpha = 0.94 \pm 0.49$), recast in the parameterization used by \cite{Wibking_2019} (hereafter Paper I).  However, we found that this parameter space was not sufficient to fit the mock galaxy catalogs we used for tests in section \ref{section:mocks}, and so we extended the parameter space to include two additional parameters, $A_{\text{conc}}$ and $R_{\text{rescale}}$, which account for differences between galaxy and halo profiles.  We also extended the minimum of the ranges of the dimensionless HOD parameters $M_0/M_1$ and $\sigma_{\log M}$ to extend to zero and increased the maximum range of the satellite galaxy mass parameter $M_1/M_\text{min}$ to 20. We determine the characteristic mass scale of halos $M_{\text{min}}$ from the desired number density of galaxies $n_{\text{gal}}$ by performing an integral over the mass function tabulated from a given simulation we wish to populate with galaxies This means that the halo occupation $\langle N_{\text{gal}}|M_{\text{halo}} \rangle$ is not fully specified by the HOD parameters alone, and must be emulated as a function of the full set of parameters in Table \ref{table:parameters}.

The $A_{\text{conc}}$ parameter is a multiplicative correction parameter to the concentrations used for the galaxy number density profile within halos, where these concentrations are taken from the $v_{\text{max}}$-based concentrations \citep{Klypin_2011} determined for each dark matter halo within the simulations by the \textsc{Rockstar} halo finding code \citep{Behroozi_2013}. As shown in Paper I, this parameter is essentially equivalent on the scales considered in this work to varying the power-law slope of the NFW profile, but it has the advantage that its convolution with itself can be written in closed form in configuration space (\citealt{Zheng_2007}; see also Appendix \ref{appendix:halo_model}). We replace the power-law slope variation parameter $\Delta \gamma$ with $A_{\text{conc}}$ in this paper.

The parameter $R_{\text{rescale}}$ is a multiplicative correction factor to the halo radii used for the galaxy number density profiles, where the halo radii are the virial radii likewise determined for each halo by \textsc{Rockstar}.  While we only have halo catalogs for spherical overdensity halos determined by the virial overdensity criterion of \cite{Bryan_1998}, and thus cannot marginalize over halo definition, marginalizing over the $R_{\text{rescale}}$ parameter allows us to take into account the main effect of varying halo definitions, namely the dependence of halo radii on halo definition.  Since the extent and spatial distribution of `satellite' galaxies is uncertain (and empirically, for luminous galaxy samples, is not the same as that of the dark matter; e.g., \citealt{Watson_2010,Piscionere_2015}), we populate halos with satellite galaxies according to an NFW profile \citep{NFW_1997} with concentration and radius corrected by $A_{\text{conc}}$ and $R_{\text{rescale}}$ from the concentration and radius determined by \textsc{Rockstar} for a given halo.

We use a cosmological parameter space that approximately encompasses the union of the WMAP9 \citep{Hinshaw_2013} and Planck \citep{Planck_2016} $w$CDM posteriors, as given by the design of the AbacusCosmos simulations \citep{Garrison_2017}. The ranges of these parameters are shown in Table \ref{table:parameters} (these ranges are not uniformly sampled, but are rather sampled along the principal components of the combined WMAP and Planck posteriors). We use the $(720 \, \hMpc)^3$ set of simulation boxes run with identical phases of their initial conditions, with particle mass $\sim 1 \times 10^{10} \, \Msunh$ at the fiducial (Planck) cosmology. Since LOWZ galaxies typically live in halos of mass $\sim 10^{13} \, \Msunh$, these halos are well resolved with $\sim 10^3$ particles per halo (although it is important to resolve halos down to the mass scale $\log M_{\text{halo}} \sim (\log M_{\text{min}} - \sigma_{\log M})$ in order to avoid biases in clustering predictions; see e.g. \citealt{Sinha_2018}).

We use simulation outputs at redshift $z=0.3$, which is close to the effective redshifts of the LOWZ sample. There are two distinct effective redshifts, one for clustering and one for lensing, due to the differing line-of-sight weight functions for each signal. We compute the effective clustering redshift as 
\begin{align}
{\langle z \rangle}_{\text{clustering}} = \frac{ \int dz \, p^2(z) \, (dV_c/dz)^{-1} \,z }{ \int dz \, p^2(z) \, (dV_c/dz)^{-1} } \, ,
\end{align}
where $p(z)$ is the weighted redshift distribution of the number of galaxies $dN_g/dz$ \citep{Mandelbaum_2011}. For the LOWZ sample we use in this work, $\langle z \rangle_{\text{clustering}} \approx 0.27$.  We compute the effective lensing redshift as 
\begin{align}
{\langle z \rangle}_{\text{lensing}} = \frac{ \int dz_l \, p_l(z_l) \, w_l(z_l) \, z_l }{ \int dz_l \,  p_l(z_l) \, w_l(z_l) } \, ,
\end{align}
with
\begin{align}
w_l(z_l) = D_L^{-2} (z_l) \, (1 + z_l)^{-2} \, \int_{z_l}^{\infty} dz_s \, p_s(z_s) \, \Sigma_c^{-2} (z_l, z_s) \, ,
\label{eq:lens_weight}
\end{align}
where $D_L(z)$ is the luminosity distance at redshift $z$, $p_s(z)$ is the source redshift distribution (weighted by the inverse variance of the shape measurements), $\Sigma_c$ is the lensing critical density, and the factor $(1+z_l)^{-2}$ accounts for the fact that in comoving coordinates lower redshift lens galaxies have larger effective apertures \citep{Nakajima_2012} This window function is equivalent to that given by \cite{Singh_2019}, assuming that differences between photometric and true source galaxy redshifts are negligible. For the spectroscopic sample and lensing catalog used here, we find $\langle z \rangle_{\text{lensing}} \approx 0.24$.

Since these redshifts are both close to the AbacusCosmos simulation outputs at $z=0.3$ \citep{Garrison_2017}, we adopt $z_{\text{eff}}=0.3$ as the effective redshift for all of our emulator predictions and neglect the difference between the clustering and lensing effective redshift.  The adequacy of this effective redshift approximation for clustering is tested by fitting to mock catalogs (section \ref{section:mocks}), while the effective redshift approximation for lensing is only approximately tested by our mock catalog fits, due to the approximation we use to compute the lensing signal from the mock catalogs. However, the tests of \cite{Singh_2019}, conducted by applying the lensing weights (eq. \ref{eq:lens_weight}) to particle-galaxy pairs in simulations, indicate that computing lensing predictions at a single effective redshift is accurate to better than $1-1.5$ per cent in this redshift range.

\subsection{Sampling strategy}

We use a modified stratified sampling design for the training dataset, generating sampling designs separately for the cosmological and non-cosmological parameters.  The sampling design for the cosmological parameters, described by \cite{Garrison_2017}, is a min-max Latin hypercube (a variant of the classical Latin hypercube with additional space-filling properties described by \citealt{Heitmann_2009}) with sampling dimensions given by the principal components of the combined WMAP and Planck posteriors and $N=40$ sub-cells per dimension (therefore yielding 40 samples). For the non-cosmological parameters, we use a Latin hypercube sampling design \citep{McKay_1979} with dimensions given by the nominal parameters in Table \ref{table:parameters} and $N=400$ cells per dimension.  Once we have generated the discrete Latin hypercube, we sample uniformly within each occupied cell of the hypercube to obtain the parameter samples, yielding 400 total samples.

We then have 400 non-cosmological parameter samples to distribute among 40 simulations. We carry this out without duplicating any non-cosmological parameter samples. For each cosmological simulation (corresponding to a realization of one of the cosmological parameter samples), we assign 10 non-cosmological parameter samples to the given simulation box from the set of unassigned non-cosmological samples.  There is therefore no special relationship assumed by the sample design between the cosmological subspace of parameters and the non-cosmological subspace of parameters. We experimented with various other sampling strategies for the subspace of non-cosmological parameters, including Latin hypercubes with additional symmetry or volume-filling properties as well as uniform random samples, and found none that offered an improvement in emulator accuracy. We note that a hybrid combination of clustered and space-filling sampling designs shows promise in other application domains \citep{Zhu_2006,Zimmerman_2006}, and we may consider such designs in future work.

\begin{table}
\caption{Emulator parameters. Each non-cosmological parameter sample is assigned to a cosmological parameter sample, with 10 non-cosmological samples assigned to a given cosmology. The emulator training set therefore has a total sample size $N=400$.}
\label{table:parameters}
\begin{tabular}{llcc}
\toprule
Parameter & Sampling range & Fiducial value & Units \\
\midrule
$n_{\text{gal}}$	& [$2.5$, $3.5$] $\times 10^{-4}$ & $3 \times 10^{-4} $ &  $(\hMpc)^{-3}$ \\
$\sigma_{\log M}$	& [0.01,	0.8] & 0.4 	& dimensionless \\
$M_0/M_1$			& [0.0,		0.4] & 0.1 	& dimensionless \\
$M_1/M_{\text{min}}$& [7.5,		20]  & 8.5 & dimensionless \\
$\alpha$			& [0.6,		1.5] & 1.0 	& dimensionless \\
$A_{\text{conc}}$	& [0.5,		3.0] & 1.0 	& dimensionless \\
$R_{\text{rescale}}$& [0.5,		2.0] & 1.0 	& dimensionless \\
\midrule
$\sigma_8$			& [0.65,	1.0]  	& 0.830      & dimensionless \\
$\Omega_{\text{CDM}} h^2$		& [0.1045, 0.1322] 	& 0.1199	  & dimensionless \\
$\Omega_b h^2$		& [0.0209, 0.0235] 	& 0.02222	  & dimensionless \\
$H_0$				& [61.567, 74.793] 	& 67.26		  & km s$^{-1}$ Mpc$^{-1}$ \\
$n_s$				& [0.9300, 0.9898] 	& 0.9652		  & dimensionless \\
$w_0$				& [-1.370, -0.655] 	& -1.0		  & dimensionless \\
$N_\text{eff}$		& ---  		& 3.046		  & dimensionless \\
\midrule
$z_{\text{eff}}$		& ---		& 0.300			& dimensionless \\
\bottomrule
\end{tabular}
\end{table}

\subsection{Emulated quantities}

\subsubsection{Projected galaxy correlation function}
\label{section:wp_ratio}
We emulate the ratio of the projected galaxy-galaxy clustering $w_p$ relative to its analytic halo model prediction:
\begin{align}
\text{ratio} \, w_p(r_p; \v{p}) = \frac{w_{p,\text{sim}}(r_p; \v{p})}{w_{p,\text{analytic}}(r_p; \v{p})} \, ,
\end{align}
where $\v{p}$ is the vector of parameters described in Table \ref{table:parameters}.
We compute $w_{p,\text{sim}}(r_p)$ via a projection integral over the real-space $\xi_{gg}$ computed from the simulations:
\begin{align}
w_p(r_p) = \int_{r_p}^{\Pi_{\text{max}}} \xigg \left( \sqrt{r_p^2 + \Pi^2} \right) \, d\Pi \, ,
\end{align}
where we choose $\Pi_{\text{max}} = 100 \, \hMpc$ and use a piecewise linear integration scheme to minimize finite bin-size effects (see Appendix \ref{appendix:projection_integrals}). This quantity is averaged over 20 stochastic realizations of the halo occupation in order to reduce noise; for a given set of parameters $\v{p}$, the mean and variance of these realizations is used as the input datapoint to the Gaussian process model (see Appendix \ref{appendix:GP}).  The analytic $w_p(r_p)$ is computed by the same projection integral over the analytic $\xigg$.

We correct for residual RSD effects on $w_p$ (which are $\approx 15$ per cent at the largest scales considered in this work) by computing the ratio between $w_p$ computed with and without the \cite{Kaiser_1987} model for redshift-space distortions on the galaxy power spectrum \citep{vdBosch_2013}:
\begin{align}
w_{\text{p,rsd}}(r_p) = \frac{1}{2\pi^2} \int_{0}^{\infty} dk_z \int_{0}^{\infty} dk_{\perp} P_{\text{gg}}(k) \left( 1 + \beta \mu^2 \right)^2  \nonumber\\
\times \cos(k_z r_{\pi}) \, J_0(k_{\perp} r_p) \, ,
\end{align}
\begin{align}
w_{\text{p,norsd}}(r_p) = \frac{1}{2\pi^2} \int_{0}^{\infty} dk_z \int_{0}^{\infty} dk_{\perp} P_{\text{gg}}(k) \nonumber\\
\times \cos(k_z r_{\pi}) \, J_0(k_{\perp} r_p) \, ,
\end{align}
where $k^2 = k_{\perp}^2 + k_z^2$, $\mu = k_z / k$, and $\beta = f(z) / b$, with $f(z)$ as the growth factor at redshift $z$ and $b$ as the large scale galaxy bias (measured from $\sqrt{P_{\text{gg}}(k)/P_{\text{mm}}(k)}$ averaged over $k \lesssim 0.1$). The galaxy power spectrum $\Pgg$ is computed via an integral over the real-space correlation function $\xigg$ measured in the simulations at the fiducial parameter values given in Table \ref{table:parameters}. We then multiply the emulated value of $w_p(r_p)$ by the ratio
\begin{align}
\text{RSD corr.}(r_p) = \frac{ w_{\text{p,rsd}}(r_p) }{ w_{\text{p,norsd}}(r_p) }
\end{align}
to obtain a scale-dependent correction factor for residual RSD effects. While this is not strictly applicable except for the fiducial parameter values, we find that this ratio is relatively insensitive to parameter changes and adopt it as a fixed correction factor for $w_p(r_p)$ at all parameter values.\footnote{Computed with traditional multidimensional cubature, the integrals involved are relatively expensive to compute. In future work, we recommend applying the \textsc{FFTLOG} method \citep{Hamilton_2000} to compute this integral (for a more complicated application, see \citealt{McEwen_2016b}).} We test the accuracy of this approximation by fitting mock catalogs with observables computed in redshift space (section \ref{section:mocks}).

\begin{figure}
  \includegraphics[width=\columnwidth]{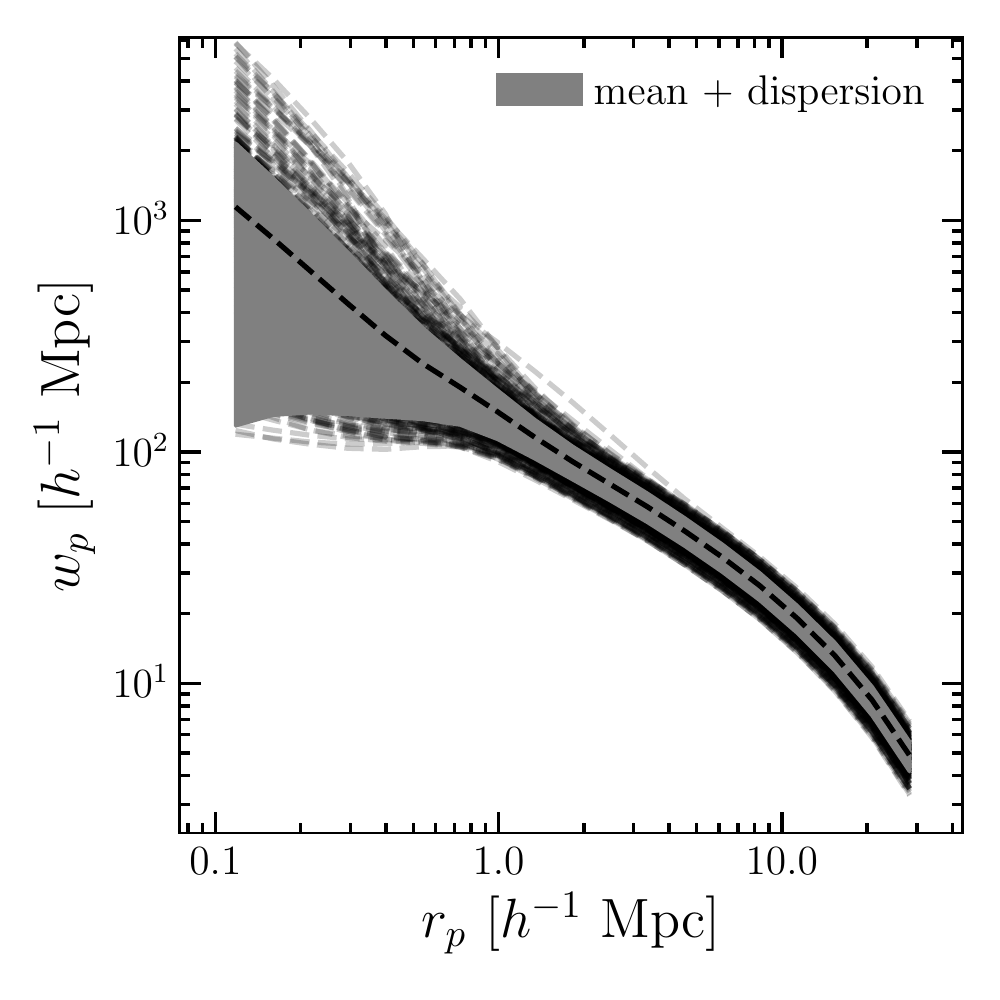}
 \caption{$w_p$ as predicted by the simulations over all $N=400$ parameter samples for all simulations used in the construction of our $w_p$ ratio emulator.}
  \label{fig:wp_raw}
\end{figure}

\begin{figure}
  \includegraphics[width=\columnwidth]{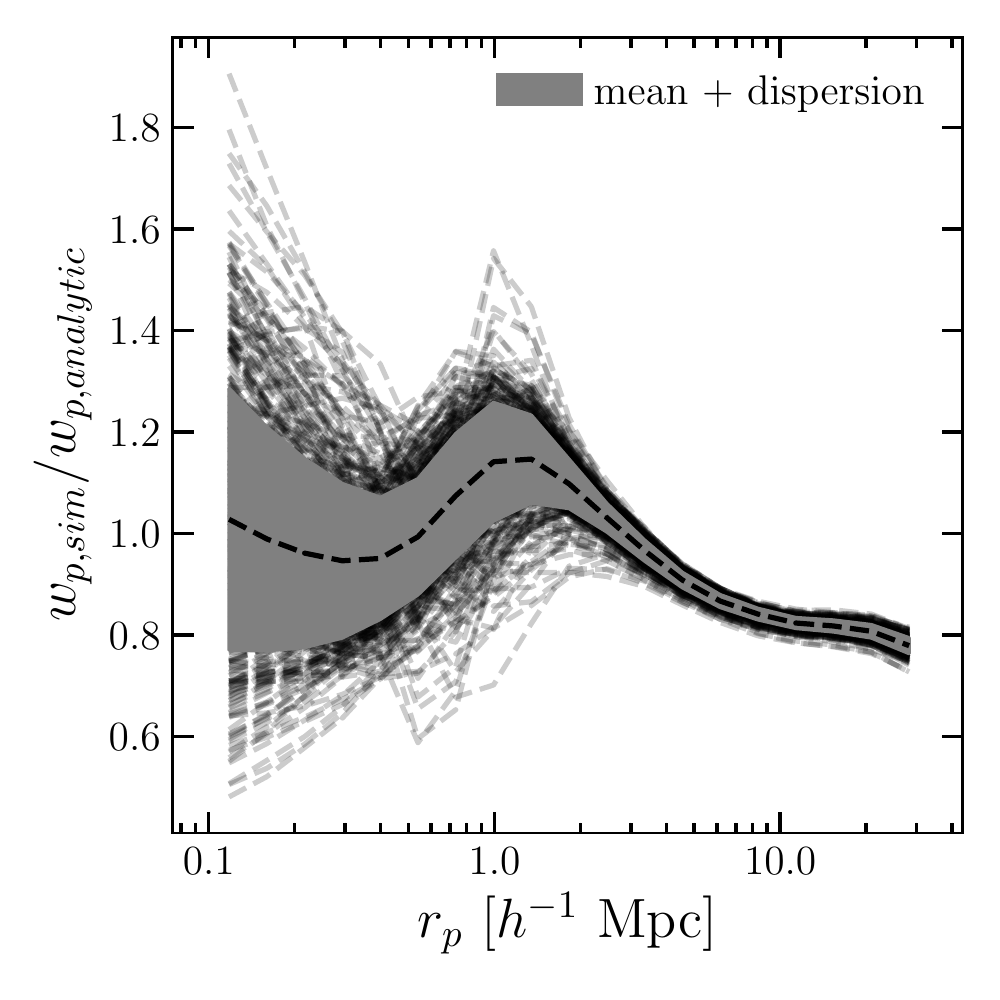}
 \caption{Ratios of $w_p$ from Figure \ref{fig:wp_raw} to $w_p$ as predicted by our analytic halo model. This quantity is used as the input to the emulator, such that the emulator only needs to learn corrections of order unity to the analytic halo model. The deviation from unity at large scales is due to an unlucky draw of the initial conditions used for the 40 cosmology-varying simulations. This is corrected by the ensemble mean correction discussed in section \ref{section:ensemble_correction}.}
  \label{fig:wp_ratios}
\end{figure}

\subsubsection{Galaxy-galaxy lensing}
For $\Delta\Sigma$, we directly emulate the observable, since it is substantially smoother (and therefore the training data have higher signal-to-noise).  $\Delta\Sigma$ is computed by integrating over $\wgm$, which in turn is computed by projecting $\xigm$ from the simulations:
\begin{align}
\Delta\Sigma(r_p) = \bar\rho \left[ \frac{4}{r_p^2} \int_{r_{p,\text{min}}}^{r_p} r \, \wgm(r) \, dr - 2 \, \wgm(r_p) \right] \, ,
\label{eq:deltasigma}
\end{align}
where
\begin{align}
\wgm(r_p) = \int_{0}^{\Pi_{\text{max}}} \xigm \left( \sqrt{r_p^2 + \Pi^2} \right) \, d\Pi \, .
\label{eq:wgm}
\end{align}
In practice both the lower limit $r_{p,\text{min}}$ and the upper limit $\Pi_{\text{max}}$ of these integrals are taken to be finite values such that the integrals converge within an acceptable precision (in our case, we use a minimum integration limit of $0.01 \, \hMpc$ and a maximum integration limit of $100 \, \hMpc$).  We likewise average this quantity over 20 realizations of the halo occupation in order to reduce noise in the training data, and use the mean and variance of these realizations as the emulator input datapoint $(y_{\text{obs,i}}, \, \sigma_{\text{obs,i}}^2)$ for a given set of parameters $\v{p}_i$.

Finally, we correct $\Delta\Sigma$ as measured in the simulations to the amplitude and radial binning an observer making the assumption of an $\Omega_m = 0.3$ cosmology would measure.  As noted in Paper I, this correction for projected distances is negligible, but the assumed-cosmology correction to the critical lensing density $\Sigma_c$ changes the degeneracy direction between $\Omega_m$ and $\sigma_8$ by $\Omega_m^{-0.1}$ (due to geometric effects on the line-of-sight lensing distance).

\begin{figure}
  \includegraphics[width=\columnwidth]{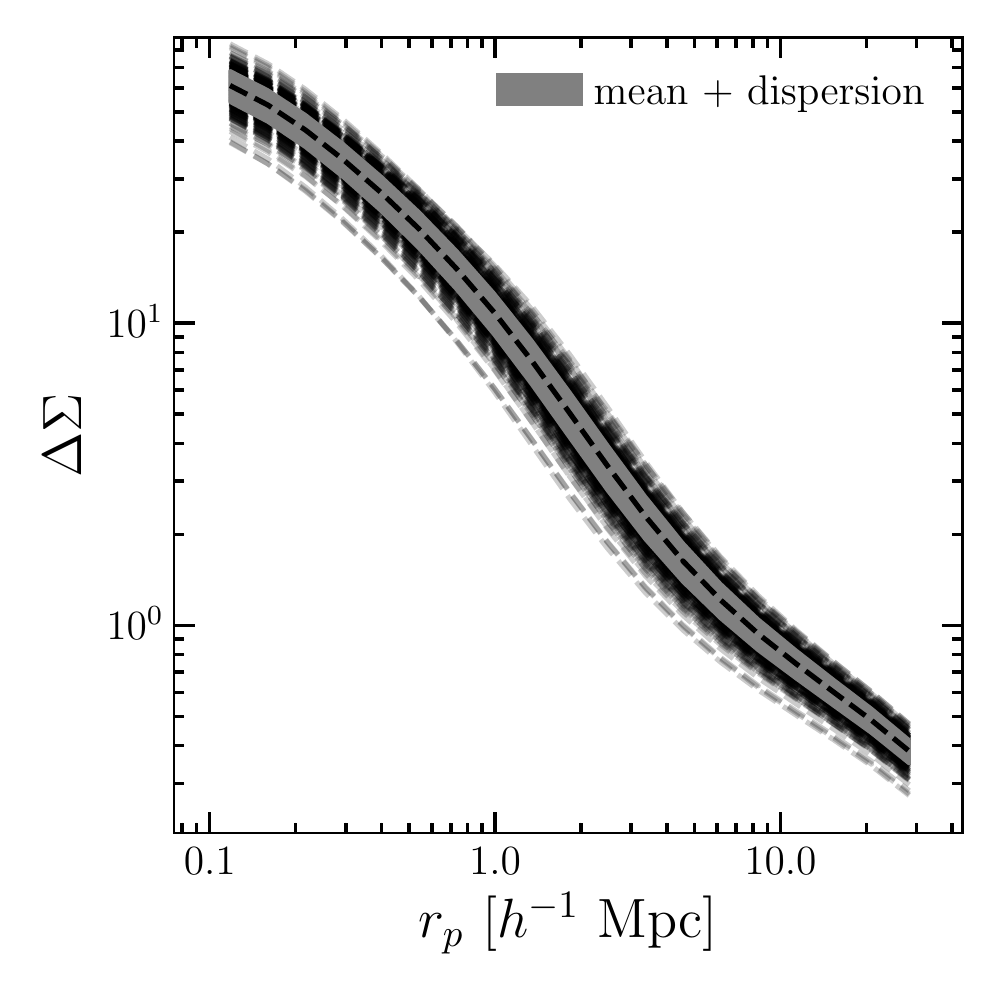}
 \caption{$\Delta\Sigma$ as predicted by the simulations over all $N=400$ parameter samples for all simulations used in the construction of our $\Delta\Sigma$ emulator.}
  \label{fig:wp_raw}
\end{figure}

\subsubsection{Ensemble mean correction}
\label{section:ensemble_correction}
Since our varying-cosmology simulations are computed with matched phases of their initial conditions (see \citealt{Garrison_2017}), we first compute the fiducial matched-phases signal from the fiducial parameters (given in Table \ref{table:parameters}) with the same phases as those used to construct the emulation training data. We then compute the ensemble mean fiducial signal by averaging over 20 additional simulations that provide realizations of the fiducial cosmology with varying phases. We then divide the ensemble mean signal by the fiducial matched-phases signal to obtain a multiplicative correction for the observables due the sample variance of our simulations.

This procedure reduces the sample variance of our training set data, the accuracy of which depends on the assumption that the derivatives of the simulated correlation functions with respect to cosmology and HOD parameters are independent of the phases of the initial conditions. Our lightcone mock tests indicate that this correction slightly reduces bias of the recovered cosmological parameters, but that our inferences of cosmological parameters are relatively unbiased even without this correction. For future datasets, a more sophisticated approach may be needed, such as the ``fixed-paired'' approach to initial conditions advocated by \cite{Angulo_2016}.

\subsubsection{Analytic halo model for $w_p$}
The analytic halo model for $w_p$ consists of a 1-halo term given by the sum of the central-satellite and satellite-satellite terms and a simple two-halo term given by the square of the large-scale bias multiplied by the linear matter correlation function computed with the `no-wiggles' fitting formula for the linear matter power spectrum of \cite{Eisenstein_1998}.  For speed, we compute the one-halo terms in configuration space, using the closed-form expression for the self-convolution of the NFW profile given by \cite{Sheth_2001} and \cite{Zheng_2007}. A complete description of the equations used in our analytic halo model is given in Appendix \ref{appendix:halo_model}.

We emulate the ratios of the observables
using Gaussian process regression \citep{Rasmussen_2006} with a squared-exponential kernel.
For each radial bin of each of the observables, we determine the
hyperparameters of the kernel by maximizing the leave-one-out cross-validation pseudo-likelihood (Appendix \ref{appendix:GP}).
This pseudo-likelihood directly minimizes the prediction error, which may be more robust to the
(arbitrary) choice of kernel function than maximizing the marginal likelihood of the Gaussian process model itself \citep{Rasmussen_2006}. We thus obtain separate Gaussian process emulators for each radial bin of each emulated quantity ($w_{p,\text{sim}} / w_{p,\text{analytic}}$ and $\Delta\Sigma$).

\subsection{Interpolation Accuracy}

\subsubsection{Cross-validation error}
We measure our emulator accuracy by using the leave-one-\emph{simulation}-out cross validation error (computed separately for each radial bin of each emulated quantity). This is defined as
\begin{align}
||\text{LOSOE}||^2 = \sum_{i}^{N} \left( \hat y_{(\text{sim}(i)), \, i} - y_i \right)^2 \, ,
\label{eq:losoe}
\end{align}
where $y_i$ is the $i$-th training datapoint value ($i=1,\dots,400$) and $\hat y_{(\text{sim}(i))}$ is the prediction of the emulator for point $i$ trained on all datapoints except for those derived from the same simulation as point $i$. This `leaves out' the training data from a given simulation when predicting the observables at the same cosmological parameters as those of the given simulation, thereby providing a conservative estimate of emulator accuracy.  Note that we fix the hyperparameters of the emulator (see Appendix \ref{appendix:GP}) when computing eq. \ref{eq:losoe}.

\begin{figure}
  \includegraphics[width=\columnwidth]{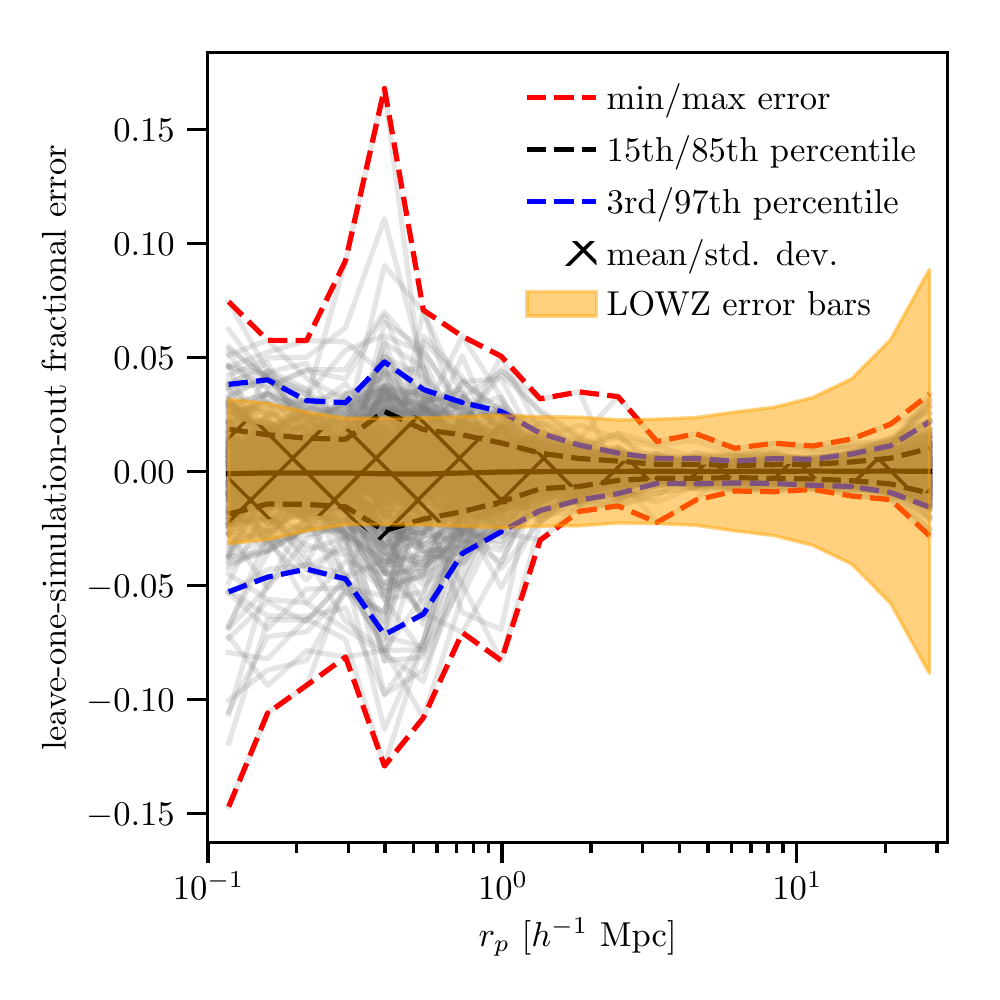}
 \caption{The error of the leave-one-simulation-out predictions of the $w_p$ ratio emulator.}
  \label{fig:emu_samples}
\end{figure}

We show emulator accuracy in Figure \ref{fig:emu_samples}. We find that emulating in logarithmic variables does not improve the emulator accuracy, so we use the linear (i.e. untransformed) variables, in contrast
to Paper I. We find that accuracy substantially increases when directly emulating the (projected) observables $w_p$ and $\Delta\Sigma$ instead of the three-dimensional correlation functions. In order to further improve accuracy, we emulate in the ratio of $w_p$ with respect to its analytic halo model prediction as described in section \ref{section:wp_ratio}.  We hypothesize that this improves emulator accuracy for $w_p$ at a fixed number of training points because the variance of the training data is reduced due to approximate prior knowledge of how the observable should respond to a parameter change.

\begin{figure}
  \includegraphics[width=\columnwidth]{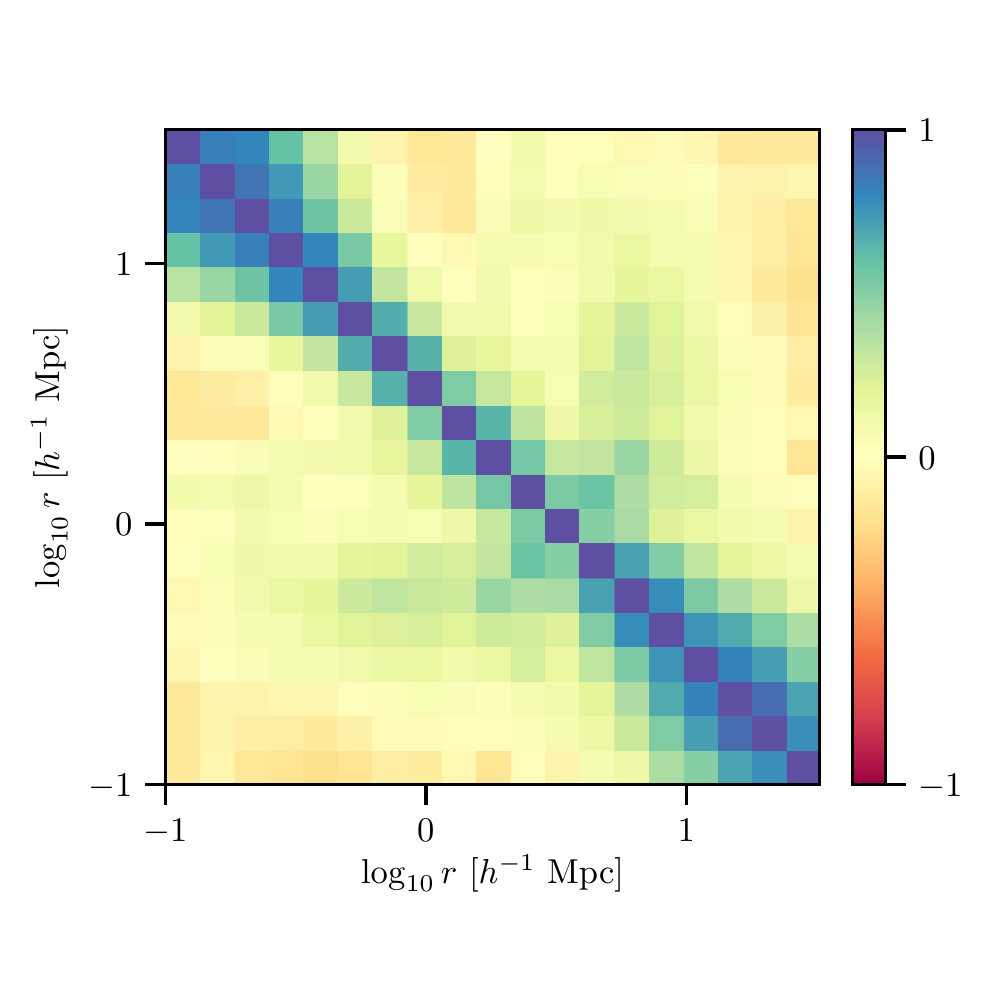}
 \caption{The correlation matrix of the leave-one-simulation-out cross-validation accuracy of the emulator between different $r_p$ bins.}
  \label{fig:emu_correlation}
\end{figure}

We also compute the covariance matrix $C$ for the leave-one-simulation-out prediction errors of our emulators across different $r_p$ bins. While one may naively expect that the errors in predicting the values of the correlation function would not be correlated between bins because we train separate emulators for each bin, this is not the case. In Figure \ref{fig:emu_correlation} we show the correlation matrix of the prediction errors for $w_p$, defined as
\begin{align}
\text{corr}_{ij} = C_{ij} / \sqrt{C_{ii} C_{jj}} \, .
\end{align}
We find that nearest-neighbor bins have highly correlated prediction errors, with a sharp drop-off in correlations for more distant bins, with the exception of bins $\gtrsim 3$ \hMpc, which are all moderately correlated with each other.  We emphasize that although the underlying reason for the correlated prediction errors is due to correlated uncertainties in the correlation function itself, the resulting pattern of correlated errors of the emulators is not identical to the form of statistical correlations in $w_p$ itself.

\subsubsection{Emulator efficiency}
Since we have virtually eliminated the noise contributions from sample variance and from
the stochasticity of the HOD (via the averaging over multiple populations described in section \ref{section:wp_ratio}), we can construct an emulator of a given accuracy
using a much smaller training set than other emulators (e.g. \citealt{Zhai_2019}, who used 2000 samples to train their GP-based emulator). This directly translates into two orders of magnitude (a factor of $[2000/400]^3 = 125$) smaller computational requirements for the hyperparameter optimization, since we gain efficiency by a factor $N^3$ in the kernel matrix inversion (eq. \ref{eq:gp_inverse}).

\section{Covariance matrix}
\label{section:covariance}

We use a combination of simulation-based and analytic covariance matrices for our mock cosmological analyses in section \ref{section:mocks}.  For $w_p$, we use a simulation-based covariance to capture the non-Gaussian part of the covariance (excess variance compared to the variance of a Gaussian field with the nonlinear galaxy power spectrum), which we find is important on 1-halo scales.  For $\Delta\Sigma$, we use an analytic covariance matrix to capture the Gaussian part of the covariance (the variance purely due to a Gaussian field with the nonlinear galaxy power spectrum), including shape noise and large-scale-structure noise.  From simulations, we find that the non-Gaussian part of the $\Delta\Sigma$ covariance is negligible, and we therefore neglect it in our analysis.\footnote{The non-Gaussian $\Delta\Sigma$ covariance would not be neglible for cluster mass halos with low shape noise weak lensing data \citep{Wu_2019}.} Since the $\Delta\Sigma$ covariance matrix is dominated by shape noise, we find that the cross-covariance between $w_p$ and $\Delta\Sigma$ is negligibly small and so we do not include it in our analysis. For completeness, we derive the Gaussian part of the cross-covariance between $w_p$ and $\Delta\Sigma$ in Appendix \ref{appendix:covariance}.

\subsection{Simulation-based covariance for $w_p$}

Due to the presence of a large non-Gaussian component on 1-halo scales, we use bootstrap-estimated simulation covariances to estimate the $w_p$ covariance matrix, populating simulation boxes with galaxies according to the fiducial HOD parameters shown in Table \ref{table:parameters}. We use 20 volumes of the Planck cosmology with varying phases from the ($1100$ \hMpc)$^3$ boxes of \cite{Garrison_2017} and divide each volume into $5 \times 5$ subvolumes which tesselate the $x$-$y$ plane in projection, for a total of $20 \times 5 \times 5 = 500$ subvolumes. We compute $w_p$ in projection within each subvolume, and compute 500 bootstrap resamples by choosing 500 subvolumes with replacement and averaging $w_p$ for each resample.  Our estimate for the covariance of $w_p$ is then the covariance among the bootstrap resamples rescaled to the effective volume $V_{\text{eff}}$ of the LOWZ galaxy sample.

\begin{figure}
  \includegraphics[width=\columnwidth]{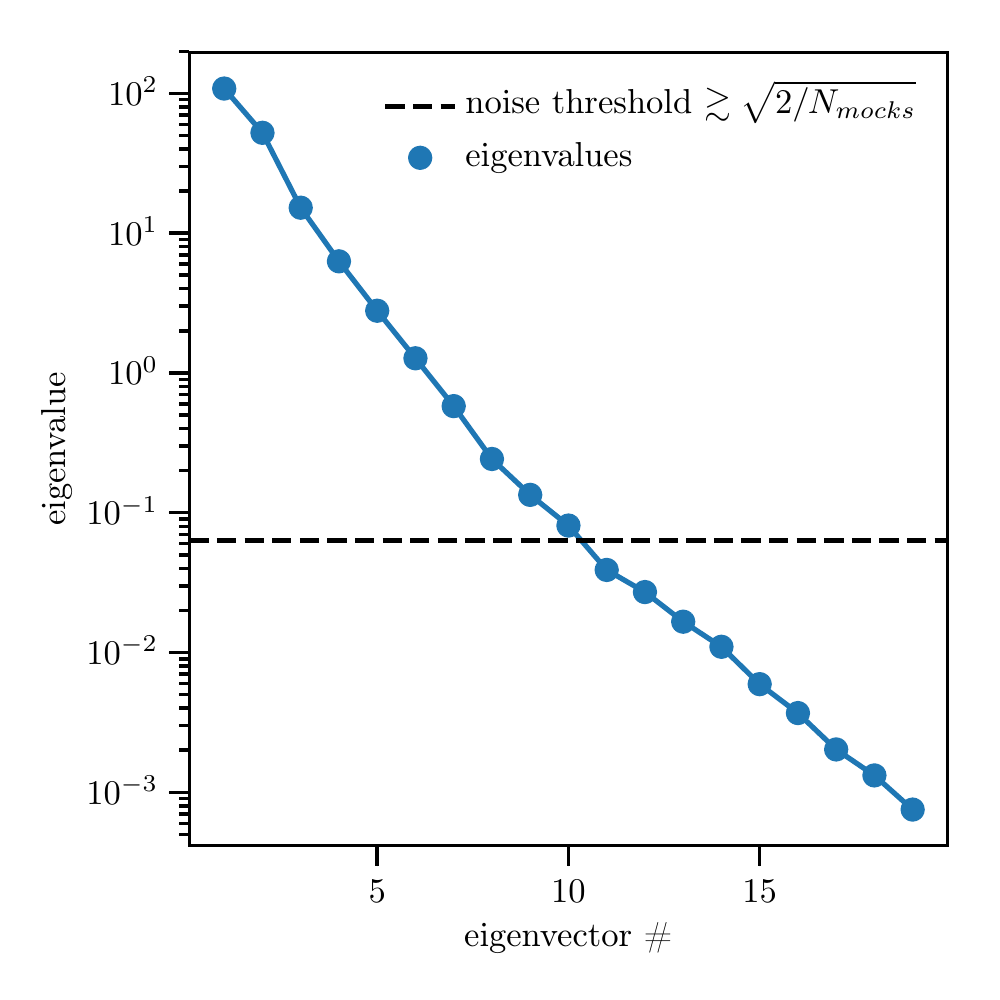}
 \caption{The eigenvalues of the covariance matrix for $w_p$ computed via $N_{\text{subvol}}=500$ simulation subvolumes.}
  \label{fig:eigenvalues}
\end{figure}

Finally, we compute the eigenvalues and eigenvectors of this covariance matrix, in order to check for noisy modes that may bias the inverse covariance matrix. A common noise threshold is
\begin{align}
\lambda_i \gtrsim \sqrt{ \frac{2}{N_{\text{subvol}}} } \, ,
\label{eq:eigenvalue_noise}
\end{align}
where $\lambda_i$ is a given eigenvalue of the covariance matrix and $N_{\text{subvol}}=500$ is the number of semi-independent subvolumes used to estimate the covariance (e.g., \citealt{Gaztanaga_2005,Sinha_2018,Zhai_2019}). While it is the most conservative choice to eliminate any potentially noisy modes when computing the inverse covariance matrix by setting the eigenvalues below the noise threshold (eq. \ref{eq:eigenvalue_noise}) to zero (then taking the reciprocal of the non-zero eigenvalues to yield the inverse covariance matrix), we did not do so in our analysis. We show the eigenvalues in Figure \ref{fig:eigenvalues} in order to illustrate that no modes are particularly noisy and therefore any noise bias in the inverse covariance matrix should be minimal.

\subsection{Analytic covariance for $\Delta\Sigma$}

We use a similar form for the Gaussian covariance for $\Delta\Sigma$ as in Paper I, but we use a slightly more accurate treatment of the large-scale-structure noise contribution (equivalent to eq. A36 of \citealt{Singh_2016}). As we derive in Appendix \ref{appendix:covariance}, the full (Limber-approximate) form of this covariance is
\begin{align}
\label{eq:ds_cov}
 \text{Cov}(&\Delta\Sigma(r_i), \, \Delta\Sigma(r_j)) = \frac{\Sigma_c^2}{V_s} \int_0^{\infty} \frac{k \, dk}{2\pi} \, \bar J_2(k r_i) \bar J_2(k r_j) \nonumber\\
	 &\times \left[ 
           \left( \Pgg(k) + \frac{1}{n_g} \right)
        \, \left( \, P_{\gamma \gamma}^{2D}(k) + \frac{\sigma_{\gamma}^2}{\Sigma_s} \right)
        +  \left( P_{\text{g} \gamma}^{2D}(k) \right)^2 \right]
\end{align}
where
\begin{equation}
    P_{\gamma \gamma}^{2D} = \int_0^{\chi_s} d\chi \, \left( \frac{\bar \rho}{\Sigma_c(\chi, \chi_s)} \right)^2 \, \Pmm\left(k \, \frac{\chi_l}{\chi}\right) \, ,
\end{equation}
\begin{equation}
    (P_{\text{g} \gamma}^{2D})^2 \approx \Pi_{\text{lens}} \left( \frac{\bar\rho}{\Sigma_c(\chi_l, \chi_s)} \right)^2 \, \Pgm^2(k) \, .
\end{equation}
Here $\Pi_{\text{lens}}$ is the effective line-of-sight depth of the squared lensing weight function, defined as
\begin{equation}
    \Pi_{\text{lens}} \equiv \int_{0}^{\chi_s} d\chi \, \left( \frac{\Sigma_c(\chi_l, \chi_s)}{\Sigma_c(\chi, \chi_s)} \right)^2 \, ,
\end{equation}
and $\Sigma_c$ is the lensing critical surface density, $\Sigma_s$ is the projected surface density of source galaxies (in units $[\hMpc]^{-2}$), $\sigma_\gamma$ is the shape noise per galaxy, $\Pgg$ is the 3D galaxy power spectrum, $\Pmm$ is the 3D matter power spectrum, $\Pgm$ is the 3D galaxy-matter cross spectrum, $\chi_l$ is the effective distance of the lens galaxy population, $\chi_s$ is the effective distance of the source galaxy population, and $\bar J_2$ denotes the bin-averaged Bessel function of order 2. The galaxy power spectra are computed from integrals over the real-space galaxy correlation functions computed at the fiducial parameters (Table \ref{table:parameters}). We note that it is necessary to average the Bessel functions over the bin widths before computing the integral (eq. \ref{eq:ds_cov}) in order to avoid divergence (see Appendix \ref{appendix:covariance} for details).

\subsection{Comparison to survey jackknife}

We find that our simulation-based covariance for $w_p$ agrees in magnitude almost perfectly with a jackknife covariance produced from the LOWZ mocks used in section \ref{section:mocks}.

The $\Delta\Sigma$ covariance cannot be directly compared to the mocks, because raytraced lensing was not done for the mocks and so any estimates from the mocks do not include shape noise or any line-of-sight large scale structure noise.  Instead, we compare the amplitude of the jackknife covariance from the LOWZ data to that of our analytic covariance, finding that the analytic covariance (at our fiducial parameter choices) has error bars that are $10-15$ per cent larger than that of the data jackknife, which we deem to be acceptable agreement given the difference of methodology.

\subsection{Point-mass marginalization}

 Since our simulations strictly only provide a prediction for $\Delta\Sigma$ up to a constant in enclosed projected mass (due to unresolved substructure at very small scales), it is necessary to include a point mass term in our cosmological analyses. This term can also absorb baryonic physics effects (such as dissipation and feedback) that are not represented in our $N$-body calculations. We use the Sherman-Morrison matrix identity in the form
\begin{align}
\tilde C^{-1} = (C + \sigma^2 vv^T)^{-1} = C^{-1} - \frac{ \sigma^2 C^{-1} v v^T C^{-1}}{1 + \sigma^2 v^T C^{-1} v} \, ,
\end{align}
where $\v{v}$ is a column vector with values $[r_{p,0}^{-2} \, , r_{p,1}^{-2} \, , ... \, , r_{p,N}^{-2}]$, and take the limit
\begin{align}
\tilde C^{-1} = \lim_{\sigma^2 \rightarrow \infty} (C + \sigma^2 vv^T)^{-1} = C^{-1} - \frac{ C^{-1} v v^T C^{-1}}{v^T C^{-1} v} \, ,
\end{align}
in order to modify our inverse covariance matrix in a way that analytically marginalizes the contribution from a point mass (e.g., \citealt{Maccrann_2019}).  The modified inverse covariance matrix $\tilde C^{-1}$ is then used in our likelihood function as such:
\begin{align}
\ln \mathcal{L} = -\frac{1}{2} \Delta y^{T} \, \tilde C^{-1} \, \Delta y - \frac{1}{2} \ln \det \tilde C - \frac{1}{2} N_{\text{obs}} \ln 2\pi \, .
\label{eq:likelihood}
\end{align}
This procedure is equivalent to explicitly marginalizing a point mass term of the form:
\begin{align}
\Delta\Sigma(r_p) = \Delta \tilde\Sigma(r_p) + \left( \frac{1}{r_p} \right)^2 \frac{M_0}{\pi} \, ,
\end{align}
with an infinitely wide Gaussian prior on the amplitude of $M_0$.  We favor this procedure over explicitly adding a point mass parameter, since no additional computational expense is added to the posterior sampling compared to neglecting the point mass term.

Adopting a point mass is equivalent to marginalizing over the uncertainty in the un-modeled inner region of the galaxy-mass cross correlation:
\begin{align}
\Delta\Sigma(r_p) &= \bar \Sigma(r_p) - \Sigma(r_p) \, , \nonumber \\
&= \bar\rho \left[ \frac{4}{r_p^2} \int_{r_{\text{p,min}}}^{r_p} r \, \wgm(r) \, dr - 2 \, \wgm(r_p) \right] \nonumber \\
&+ \bar \rho \left[ \frac{4}{r_p^2} \int_0^{r_{\text{p,min}}} r \, \wgm(r) \, dr \right] \, , \nonumber \\
&= \Delta \tilde \Sigma(r_p) + \bar \rho \left[ \frac{4}{r_p^2} \int_0^{r_{\text{p,min}}} r \, \wgm(r) \, dr \right] \, , \nonumber \\
&= \Delta \tilde \Sigma(r_p) + \left( \frac{r_{\text{p,min}}}{r_p} \right)^2 \bar \Sigma(r_{\text{p,min}}) \, ,
\end{align}
where we can identify the enclosed mean projected mass $\pi (r_{\text{p,min}})^2 \, \bar \Sigma(r_{\text{p,min}})$ with the point mass $M_0$.  A related (but not equivalent) approach is that of \cite{Baldauf_2010}, who construct an estimator that effectively subtracts the point mass term from the observations at the cost of increased noise. This approach is used by \cite{Singh_2018}. Our approach does not increase the noise of the lensing signal itself but instead requires marginalization over the amplitude of the point mass.

Since our procedure makes the updated covariance matrix formally singular (i.e. $\det \tilde C = 0$), we likewise update the $\frac{1}{2} \ln \det \tilde C$ term of eq. \ref{eq:likelihood} according to eq. 10 of \cite{Bridle_2002} in order to correctly normalize the likelihood and therefore obtain a correct value of the Bayesian evidence integral
\begin{align}
\mathcal{Z} = \int \mathcal{L}(\v{p}) \, p(\v{p}) \, d \v{p} \, ,
\label{eq:evidence}
\end{align}
where $\v{p}$ is the vector of parameters, $p$ is the prior (defined in Table \ref{table:priors}), $\mathcal{L}$ is the likelihood (implicitly dependent on the observed data; eq. \ref{eq:likelihood}), and $\mathcal{Z}$ is the evidence, or normalization constant for the posterior distribution.

\section{Cosmological analysis on mocks}
\label{section:mocks}

We test the ability of our emulator-based model to recover the correct cosmological parameters from a mock galaxy catalog produced from an independent simulation (\citealt{Klypin_2016}; run with a different $N$-body code, \textsc{GADGET-2}; \citealt{Springel_2005}) with an independent method for populating halos with galaxies (subhalo abundance matching; e.g. \citealt{Kravtsov_2004,Vale_2004}).  This is the same galaxy sample lightcone as used in \cite{Singh_2018}, produced with subhalo abundance matching (SHAM) tuned to match the number density and clustering of LOWZ galaxies in three disjoint redshift bins in the range $0.16 < z < 0.36$ (originally described in \citealt{Nuza_2013,RodriguezTorres_2016}). The lightcone was generated by populating subhalos found with \textsc{Rockstar} \citep{Behroozi_2013}, using its default halo definition corresponding to the virial mass given by the fitting function of \cite{Bryan_1998}. Because we have only a single lightcone volume, we can only test for biases of our parameter constraints at the level of the statistical uncertainty of the LOWZ data set. However, we do construct variants of the lightcone catalog to test for specific possible systematic effects as discussed below.

The projected clustering $w_p$ is computed by converting the mock catalog galaxy coordinates into redshift space comoving distances and counting galaxy pairs to tabulate the statistic $\xigg(r_p, \Pi)$ in transverse bins $r_p$ and line-of-sight bins $\Pi$ with the Landy-Szalay estimator \citep{Landy_1993},
\begin{align}
\xigg(r_p, \Pi) = \frac{DD - 2 DR + RR}{RR} \, ,
\end{align}
then integrating along the line of sight:
\begin{align}
w_p(r_p) = 2 \int_0^{\Pi_{\text{max}}} \xigg(r_p, \Pi) \, d\Pi
\end{align}
with $\Pi_{\text{max}} = 100$ \hMpc.

The galaxy-galaxy lensing signal $\Delta\Sigma$ is computed from the mock by computing the galaxy-matter correlation function
\begin{align}
\xigm(r_p, \Pi) = \frac{D_1 D_2 - D_1 R - D_2 R + RR}{RR}
\end{align}
where $D_1$ corresponds to galaxies and $D_2$ corresponds to matter particles
and integrating along the line-of-sight to obtain
\begin{align}
\wgm(r_p) = \int_{0}^{\Pi_{\text{max}}} \xigm (r_p, \Pi) \, d\Pi \, , 
\end{align}
where $\Pi_{\text{max}} = 100 \, \hMpc$, then integrating once more (using eq. \ref{eq:deltasigma}) to obtain $\Delta\Sigma$, using $r_{p,\text{min}} = 0.1 \, \hMpc$. We add an additional term to $\Delta\Sigma$ based on a power-law extrapolation of $\wgm(r_p)$ in order to account for the projected mass at scales $< r_{p,\text{min}}$:
\begin{align}
\bar \rho \, \frac{4}{r_p^2} \int_0^{r_{p,\text{min}}} r \, \wgm(r) \, dr = \, &4 \bar \rho \, \wgm( r_{p,\text{min}}) \, \left( \frac{ r_{p,\text{min}} }{ r_p } \right)^2 \, ,
\end{align}
where we assume $\wgm(r_p) \propto 1/r_p$.

This calculation does not precisely respect the true redshift weighting of the galaxy-shear signal over the range of the LOWZ sample, which has a redshift-dependent lensing kernel. A fully correct calculation requires raytraced simulations with resolved halo substructure and an explicit source galaxy population, which are not available to us.  However, the difference between the true $\Delta\Sigma$ and the approximate signal used here has been calculated (using the lensing weights in eq. \ref{eq:lens_weight} applied to simulation particle-galaxy pairs) to be $\lesssim 1-1.5$ per cent at any projected scale used here \citep{Singh_2019}.

\subsection{Test of Emulator}

For the fiducial mock, we adopt the emulator model described in section \ref{section:emulator} and the covariance matrix described in section \ref{section:covariance} and compute the Bayesian evidence of the model via nested sampling \citep{Skilling_2004} with the \textsc{MultiNest} implementation \citep{Feroz_2008}. This method integrates Eq. \ref{eq:evidence} by sampling from the prior and then rejecting samples that lie outside of successively-smaller nested likelihood contours. The \textsc{MultiNest} implemention of nested sampling approximates each likelihood contour with a set of bounding ellipsoids which attempt to fully enclose the given likelihood contour, thereby gaining efficiency compared to rejection sampling from the entire prior volume.  The sampling stops when the code estimates that the log evidence lower bound computed via nested sampling is within 0.05 of the log evidence upper bound estimated via the size of the remaining likelihood contour. The most significant failure mode of this method is to fail to sample one of the modes of a multi-modal posterior in the first iteration of the algorithm, which we attempt to guard against by checking that the posterior and evidence are stable against increasing the initial number of sampling points (so-called `live points') by a factor of 2.\footnote{We use at least 400 initial samples for all posteriors and evidence computations used here. We do not use importance nested sampling but instead use the more conservative classical \textsc{MultiNest} algorithm as implemented in \textsc{MultiNest} version 3.11.} As a byproduct of the Monte Carlo evaluation of the evidence integral, we also obtain the posterior distribution of the parameters \citep{Skilling_2004}. This method has compared favorably with Markov Chain Monte Carlo posterior estimation in the \cite{DESY1KP} cosmological analysis. 

As a conservative test of the accuracy of our emulator, we do not add any additional uncertainties to our analysis covariance matrix, with the understanding that any emulator inaccuracies (due to interpolation error or the finite volume of our simulations, or due to systematic errors in our methodology) could cause biases in parameter recovery.  In this case, a lack of identifiable bias will indicate that any emulator inaccuracies are subdominant to the statistical precision of the measurements, as we will show is the case in the following sections.

We adopt the prior ranges shown in Table \ref{table:priors}.
\begin{table}
\caption{Priors on parameters for our cosmological analyses. Units are given in Table \ref{table:parameters}.}
\label{table:priors}
\begin{tabular}{ll}
\toprule
Parameter & Prior range \\
\midrule
$n_{\text{gal}}$	& [$2.5$, $3.5$] $\times 10^{-4}$ \\
$\sigma_{\log M}$	& [0.01,	0.8] \\
$M_0/M_1$			& [0.0,		0.4] \\
$M_1/M_{\text{min}}$& [7.5,		20] \\
$\alpha$			& [0.5,		1.5]\\
$A_{\text{conc}}$	& [0.5,		3.0]\\
$R_{\text{rescale}}$& [0.5,		2.0] \\
\midrule
$\sigma_8$			& [0.65, 1.0]  \\
$\Omega_m$		& [0.26, 0.35] \\
$\Omega_b$		& [0.0394, 0.0602]  \\
$H_0$				& [61.57, 74.80]  \\
$n_s$				& [0.93, 0.9898]  \\
$w_0$				& [-1.37035, -0.6548324] \\
\midrule
$A_{\text{lensing}}$	& $\mathcal{N}(\mu=1.0, \sigma=0.06)$ \\
\bottomrule
\end{tabular}
\end{table}
We show the projections of our parameter posteriors in Figure \ref{fig:posterior_mocks} and tabulate the posterior means and 68 per cent credible intervals in the leftmost column of Table \ref{table:posterior_mocks}.  The parameter of greatest interest is the combined cosmological parameter $S_8$, defined as
\begin{align}
S_8 \equiv \left( \frac{\sigma_8}{0.8228} \right) \left( \frac{\Omega_m}{0.3107} \right)^{0.6} \, ,
\end{align}
which is chosen such that it is approximately the best-constrained combination of $\Omega_m$ and $\sigma_8$ (as determined in Paper I), and is normalized to unity for the true values of $\sigma_8$ and $\Omega_m$ used in the lightcone simulation, so as to be consistent with the definition used by \cite{Singh_2018}.\footnote{We note that this definition is \emph{not} consistent with that used by \cite{DESY1KP}.}  For the fit to the fiducial mock, we recover this parameter to within 1.7 per cent of its true value, which corresponds to approximately half of its posterior uncertainty. Because we have only a single mock realization of the LOWZ volume, we cannot test our recovery method at a precision higher than the statistical error of the LOWZ data.

Since our model is nonlinear, we cannot use the classical $\chi^2$ goodness-of-fit test to assess whether our model is consistent with the mock data.\footnote{This is because the number of degrees of freedom of a nonlinear model is not well-defined. See \cite{Andrae_2010} for a pedagogical explanation.}  Instead, we use the Bayesian posterior predictive discrepancy with a $\chi^2$-like test statistic \citep{Guttman_1967,Gelman_1996} to quantify the tension of the posterior with the mock data.  The test statistic $\Delta\chi^2$ is computed for the data by finding the minimum $\chi^2$ between the data vector and all of the posterior samples. In the limit of a large number of posterior samples, this is equivalent to the classical $\chi^2$ statistic computed from the data. However, the reference distribution differs from the classical test, and we compute the reference (i.e. expected) distribution of $\Delta\chi^2$ via a Monte Carlo procedure. For each posterior sample, we add noise by sampling with our fiducial covariance matrix, then compute the minimum $\chi^2$ between the noisy sample and all other posterior samples.\footnote{We choose a `minimum discrepancy' statistic rather than an `average discrepancy' statistic \citep{Gelman_1996} because the latter is more sensitive to the normalization of the covariance matrix, which we do not attempt (and do not need, for the purpose of parameter inference) to compute to the same accuracy as our model predictions.} By comparing the $\Delta\chi^2$ from the data with this reference distribution, we obtain a convenient measure of how consistent the data are with being drawn from the posterior distribution. We find that the posterior test statistic $\Delta\chi^2$ for the mock data is smaller than that of $68.3$ per cent of the posterior samples.  Interpreting this result as a frequentist statistical test, we conclude that there is no significant evidence against the null hypothesis that the data are drawn from the posterior. We emphasize that this test cannot give the probability that our model is correct (for which Bayesian model comparison is required), but rather only suggests that our model (considered without reference to other models) is an adequate description of the mock data given the uncertainties.

\subsection{Varying satellite fractions and incompleteness}

We vary the satellite fraction in the mock, as described by \cite{Singh_2018}, and recompute the posteriors with \textsc{MultiNest}.  For both a 15 per cent lower satellite fraction and a 15 per cent higher satellite fraction, we recover the true value of $S_8$ within $\sim 0.5 \sigma$.

We test the effect of a 15 per cent incompleteness fraction (applied uniformly to both central and satellite galaxies) by increasing the number density and then stochastically removing 15 per cent of all galaxies such that the observed number density of the mock catalog is consistent with the observed number density of LOWZ galaxies. For this mock, we recover the true value of $S_8$ within $\sim 0.5 \sigma$ (see Table \ref{table:posterior_mocks}).

\subsection{Parameterized lensing systematics}
\label{section:A_lensing}

\begin{figure*}
  \includegraphics[width=\textwidth]{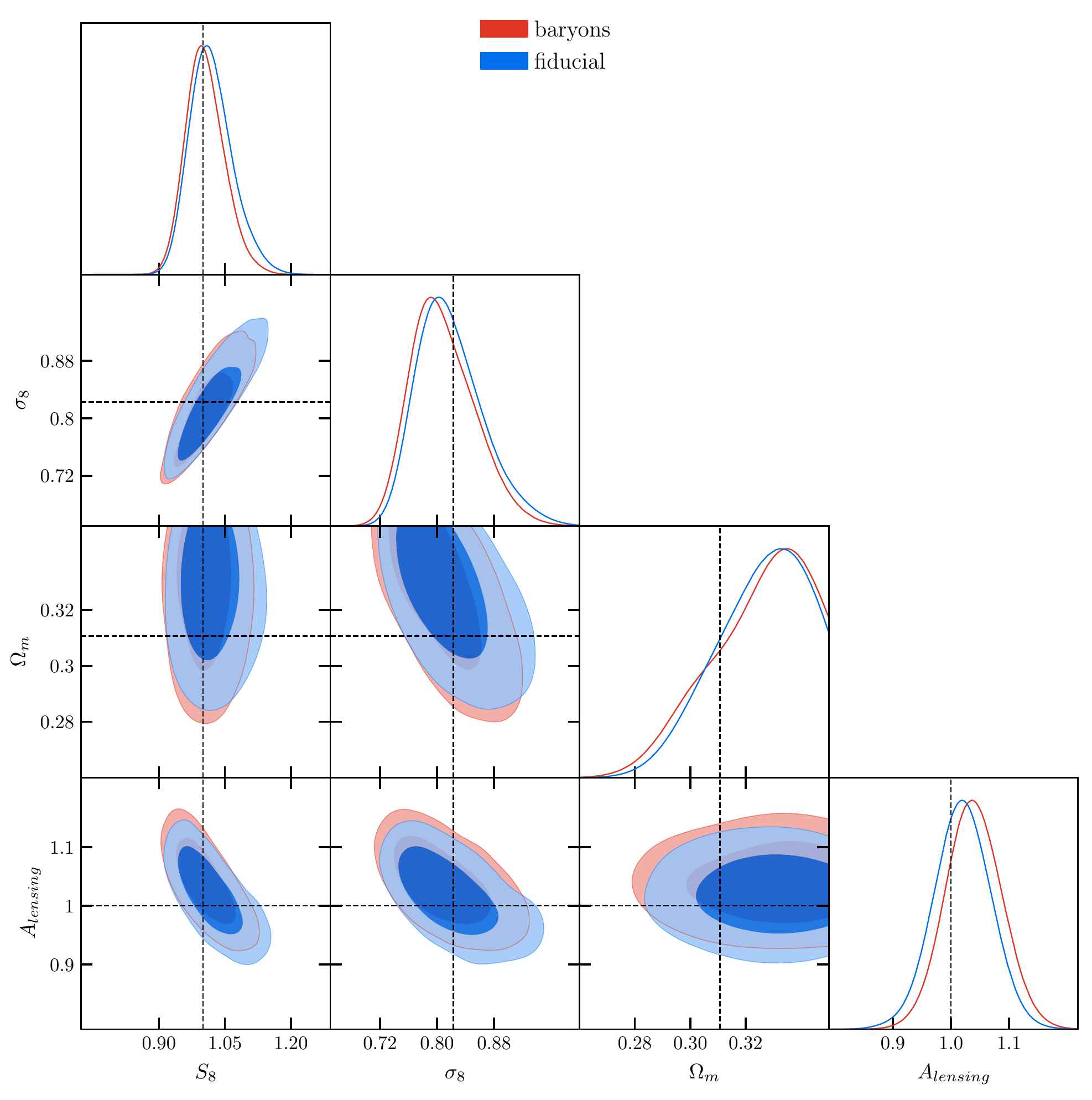}
 \caption{Posterior parameter distribution for fit of emulator to $w_p$ and $\Delta\Sigma$ measurements from the fiducial LOWZ \emph{mock} and to measurements with baryonic effects added (red contours/curves) as described in section \ref{section:baryons}. Dashed lines show the true parameter values in the mock. These fits include a nuisance parameter for lensing systematics as described in section \ref{section:A_lensing}.}
  \label{fig:posterior_mocks}
\end{figure*}

For the previous tests, we assumed that neither our emulator nor our mock data contributed any additional systematic errors, and so the resulting biases would be conservative.  However, we computed an additional mock recovery test including a multiplicative lensing calibration parameter to account for the combined effects of multiplicative shear calibration bias and photometric redshift calibration bias (which becomes multiplicative in $\Delta\Sigma$ through its multiplicative effect on the critical lensing surface density $\Sigma_c$).  The prior on this calibration parameter was set equal to $\mathcal{N}(\mu=1, \sigma=0.06)$, consistent with the calibration prior used on the analysis with LOWZ data (see following section). We did not change the mock data themselves, so the true value of this parameter for all of our lightcone test scenarios is $A_{\text{lensing}}=1.0$. In this scenario, we find the recovered $S_8$ to be within 2 per cent of the true value (within $\sim 0.3 \sigma$), and we find the recovered calibration parameter $A_{\text{lensing}}$ to be $\sim 3$ per cent lower than the true value (within $\sim 0.5 \sigma$), as shown in Figure \ref{fig:posterior_mocks}.  Adding the $A_{\text{lensing}}$ parameter increases the $S_8$ uncertainty from 0.029 to $0.048$, implying that 40 per cent of the error budget is due to systematic uncertainties in the lensing signal. We show the posterior mean values of our parameters for all of the tests of this section in Table \ref{table:posterior_mocks}.

\begin{table*}
\caption{Posterior means ($\pm 68$ per cent credible intervals) of the parameters resulting from our cosmological analyses on both lightcones (indicated with non-boldface column names) and data (indicated with boldface column names). Units are given in Table \ref{table:parameters}. Upper or lower limits indicate that the 68 per cent credible interval abuts the prior minimum or maximum value of a parameter. For $A_{\text{conc}}$ in the `baryons' column, the 68 per cent posterior interval is identical to the prior, so no values are given in the table for this entry. For entries marked `N/A', the analysis did not include $A_{\text{lensing}}$ as a free parameter. Due to the asymmetry of the marginalized posteriors, the posterior means are not necessarily identical to the posterior maxima shown in Figures \ref{fig:posterior_mocks} and \ref{fig:posterior_data}.}
\label{table:posterior_mocks}
\input{./Posteriors/lowz_allruns_posterior_table.tex}
\end{table*}

\section{Cosmological analysis on data}
\label{section:data}

We determined the design of the emulator, computed the covariance matrices, and performed all tests on the mock catalogs discussed in the previous sections prior to computing any posteriors from the data.  We formally blinded our analysis by multiplying the observed $\Delta\Sigma$ (on all scales) by a constant drawn from $\mathcal{N}(\mu=1.0, \sigma=0.06)$ and only known to one of us before unblinding.\footnote{We note that in the context of Bayesian model averaging, which we advocate in section \ref{section:conclusions}, blinding is superfluous, inasmuch as confirmation bias is caused by allowing for post-hoc model selection. In practice, however, there may be some benefit.} The width of this distribution was chosen to be moderately larger than the expected statistical precision of our posterior inference on $S_8$. From tests on our fiducial mock, we find that this constant is strongly degenerate with $\sigma_8$, so it meaningfully blinded our posterior inference of the combined parameter $S_8$, which is the only cosmological parameter with strong constraints in this analysis.

We use the measurements of LOWZ clustering and galaxy-galaxy lensing ($w_p$ and $\Delta\Sigma$) from \cite{Singh_2018}.  To summarize the data sources: for the galaxy sample, we use the Baryon Oscillation Spectroscopic Survey (BOSS) Data Release 12 (DR12) LOWZ sample \citep{Alam_2015,Reid_2016}, which has an effective area of 8,337 deg$^2$ (including NGC and SGC regions, rejecting the small part of the NGC region that had incorrect targeting for the LOWZ sample, and weighting the resulting sky area by observational completeness). We select a subsample of this catalog to include only galaxies in the redshift range $0.16 < z < 0.36$ and further reject galaxies within regions that do not pass the photometric quality cuts for shape measurements used by \cite{Singh_2017} (originally defined by \citealt{Reyes_2012}), removing an additional 8 per cent of galaxies.  The weights applied to each galaxy are the combined systematic, fiber collision, and redshift failure weights developed by the BOSS large-scale structure working group \citep{Ross_2014}. We only use scales for which the fiber collision weights provide a correction that is accurate at the $\lesssim 1$ per cent level, above approximately twice the projected scale of fiber collisions at the maximum redshift of the sample (section \ref{section:systematics}).

For the galaxy shear catalog, we use a catalog  derived from the Sloan Digital Sky Survey (SDSS) Data Release 8 (DR8) imaging \citep{Aihara_2011} with the photometric calibration method of \cite{Padmanabhan_2008}. Galaxy shapes were measured with the re-Gaussianization method \citep{Hirata_2003}, with the shear response calibrated via image simulations that included the effects of nearby neighbors \citep{Mandelbaum_2012,Mandelbaum_2017}.  Photometric redshifts of this catalog were measured by \cite{Nakajima_2012} and \cite{Reyes_2012} with the \textsc{ZEBRA} code \citep{Feldmann_2006} and tested with clustering redshifts by \cite{Singh_2019}. For a more complete description, we refer the reader to \cite{Singh_2019}. We allow for lensing systematic uncertainties by including the $A_{\text{lensing}}$ parameter with a 6 per cent prior.

\begin{figure*}
  \includegraphics[width=\textwidth]{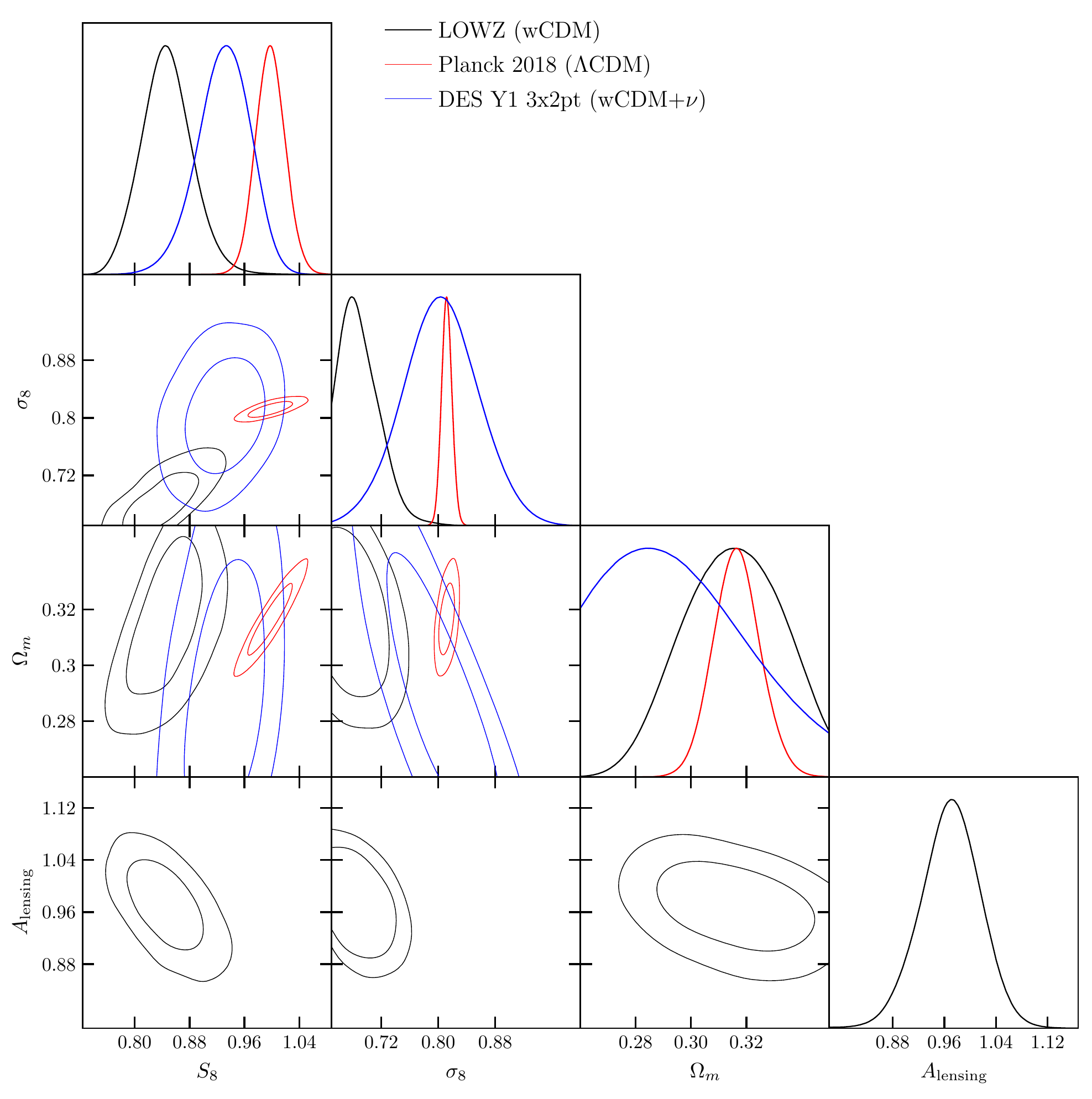}
  \caption{Posterior parameter distribution for fiducial fit of emulator to LOWZ data (black contours/lines). The red contours/lines show the posterior parameter space for the fiducial $\Lambda$CDM analysis of Planck Collaboration et al. (2018). The blue contours/lines show the posterior parameter space for the fiducial analysis of DES Year 1 galaxy clustering, galaxy-galaxy lensing, and cosmic shear data, varying the equation of state of dark energy and the sum of neutrino masses (DES Collaboration et al. 2017).}
  \label{fig:posterior_data}
\end{figure*}

\begin{figure}
  \includegraphics[width=\columnwidth]{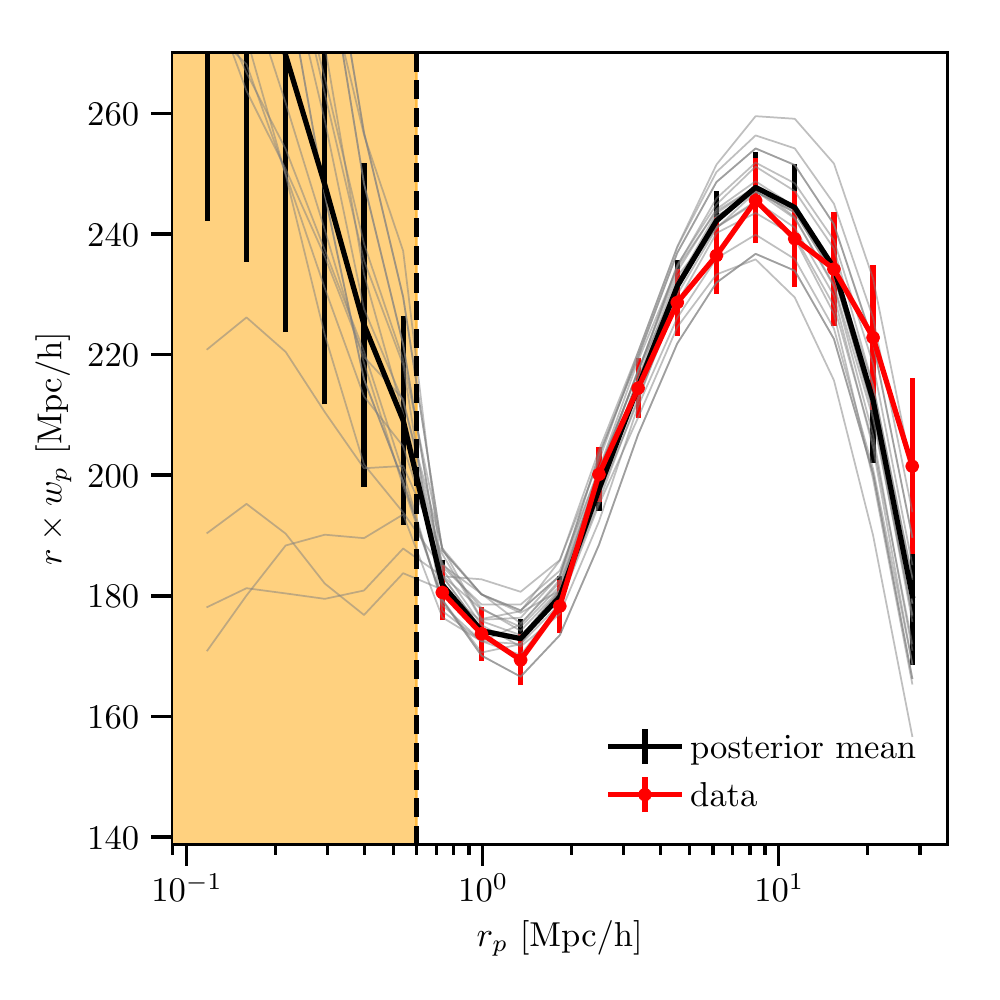}
 \caption{Posterior predictive distribution for $w_p$ when fit to LOWZ data. The data is shown as the solid red line with circles plotted for individual data points. The solid black line is the mean of the $w_p$ signal computed from the posterior samples, with the error bars indicating the standard deviation of $w_p$ computed from the posterior samples. The gray lines are individual model predictions for $w_p$ drawn at random from the set of posterior samples. While there is good agreement between the data and the posterior overall, there is a hint of tension on scales $\gtrsim 5 \, \hMpc$.}
  \label{fig:ppd_wp_data}
\end{figure}

\begin{figure}
  \includegraphics[width=\columnwidth]{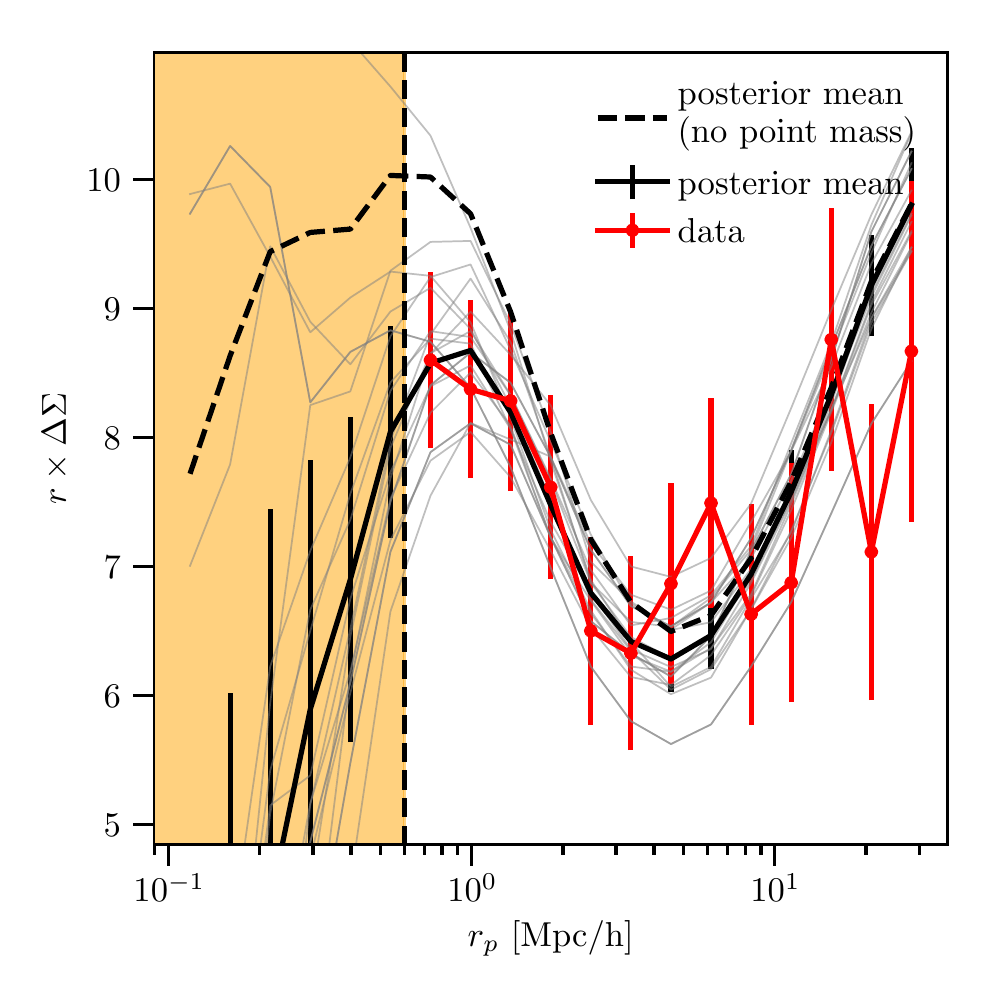}
 \caption{Posterior predictive distribution for $\Delta\Sigma$ when fit to LOWZ data. The data is shown as the solid red line with circles plotted for individual data points. The solid black line is the mean of the $w_p$ signal computed from the posterior samples, with the error bars indicating the standard deviation of $w_p$ computed from the posterior samples. The gray lines are individual model predictions for $w_p$ drawn at random from the set of posterior samples. The dashed black line is the posterior mean computed without a point mass term, illustrating that the data favor a negative value of the point mass term. There is good agreement between the data and the posterior when the point mass term is included, with a slight hint of possible tension on scales $\gtrsim 10 \, \hMpc$.}
  \label{fig:ppd_ds_data}
\end{figure}

In Figure \ref{fig:posterior_data}, we show the posterior parameter contours inferred from our fiducial cosmological analysis for the parameters $S_8$, $\sigma_8$, $\Omega_m$, and $A_{\text{lensing}}$. However, we marginalize over all parameters shown in Table \ref{table:priors}, including all $w$CDM cosmological parameters (except $N_{\text{eff}}$). We see that $S_8$ is degenerate with $A_{\text{lensing}}$, as expected from our tests on mocks, and that this degeneracy substantially increases the uncertainty of our final measurement on $S_8$. The parameter $S_8$, while almost perfectly decorrelated with $\Omega_m$ in our tests on mocks (Figure \ref{fig:posterior_mocks}), is not exactly the best-constrained combination of $\sigma_8$ and $\Omega_m$ for the data, since $\Omega_m$ is not fully decorrelated with it. Our fiducial measurement of $S_8$ is $0.847 \pm 0.037$, a 4.4 per cent measurement.  With the results of \cite{Planck_2018} giving $S_8 = 1.00 \pm 0.02$ for their fiducial flat $\Lambda$CDM model, this represents a $3.5\sigma$ tension with \emph{Planck} data. We show the posterior predictive distribution for $w_p$ in Figure \ref{fig:ppd_wp_data} and for $\Delta\Sigma$ in Figure \ref{fig:ppd_ds_data}. While there is very good agreement between the data and the model overall, there may be a hint of tension on scales $\gtrsim 5-10 \, \hMpc$.  In Figure \ref{fig:ppd_ds_data}, we show a model computed without a point mass term in addition to our fiducial model with a point mass term, illustrating that the data favor a negative value of the point mass term and that this term is important to obtain a model that is in good agreement with the data on small scales.

We additionally compute the posterior resulting from only using scales in the 2-halo regime ($\gtrsim 2 \, \hMpc$). The marginalized parameter constraints for this analysis are shown in the rightmost column of Table \ref{table:posterior_mocks}.  We find that the posterior mean of $S_8$ is consistent in this analysis with the posterior mean obtained from our fiducial analysis that includes two-point information from the 1-halo regime, but with a larger, 6 per cent uncertainty, $S_8 = 0.85 \pm 0.05$ (quoting the standard deviation instead of the 68 per cent credible interval as used in Table \ref{table:posterior_mocks}). Incorporating the uncertainty in the Planck analysis, this is a $2.6 \sigma$ tension between our results and the fiducial Planck cosmological analysis.  While there are small ($\sim 1 \sigma$) parameter shifts for some of the HOD parameters compared to our fiducial analysis, the agreement of the cosmological parameter values between these analyses suggests that any possible tension between small and large scales does not affect the cosmological parameters of interest.

In Paper I, we predicted that including 1-halo information down to $\sim 0.5 \, \hMpc$, compared with using only 2-halo information, could improve constraints on $S_8$ by a factor of $\sim 2$. This is not achieved in our LOWZ analysis because much of the total uncertainty in both cases comes from the weak lensing systematics parameterized by $A_{\text{lensing}}$. To explicitly verify this point, we analyzed the data on the fiducial set of scales ($0.6 \lesssim r_p \lesssim 30$) with the parameter $A_{\text{lensing}}$ fixed to 1 and without including the point mass term, yielding $S_8 = 0.803 \pm 0.023$, a 2.8 per cent measurement, and additionally with $A_{\text{lensing}}$ fixed to 1 but with the point mass term, yielding $S_8 = 0.831 \pm 0.029$, a 3.5 per cent measurement (in contrast to 4.4 per cent precision when both including a point mass term and marginalizing over $A_{\text{lensing}}$). We do not consider these modified analyses as part of our science results, except to illustrate that if the $\Delta\Sigma$ signal could be fully modeled without additional nuisance parameters, then precision in good agreement with our predictions from Paper I can be obtained (where we predicted 2 per cent uncertainty on $S_8$ in the case where the only cosmological parameters marginalized over were $\Omega_m$ and $\sigma_8$). Current weak lensing surveys (e.g. \citealt{DESY1KP}) aim to reduce photometric redshift zero-point and shear calibration uncertainties to the percent or sub-percent level so that they can take full advantage of their smaller statistical errors.

\begin{figure}
  \includegraphics[width=\columnwidth]{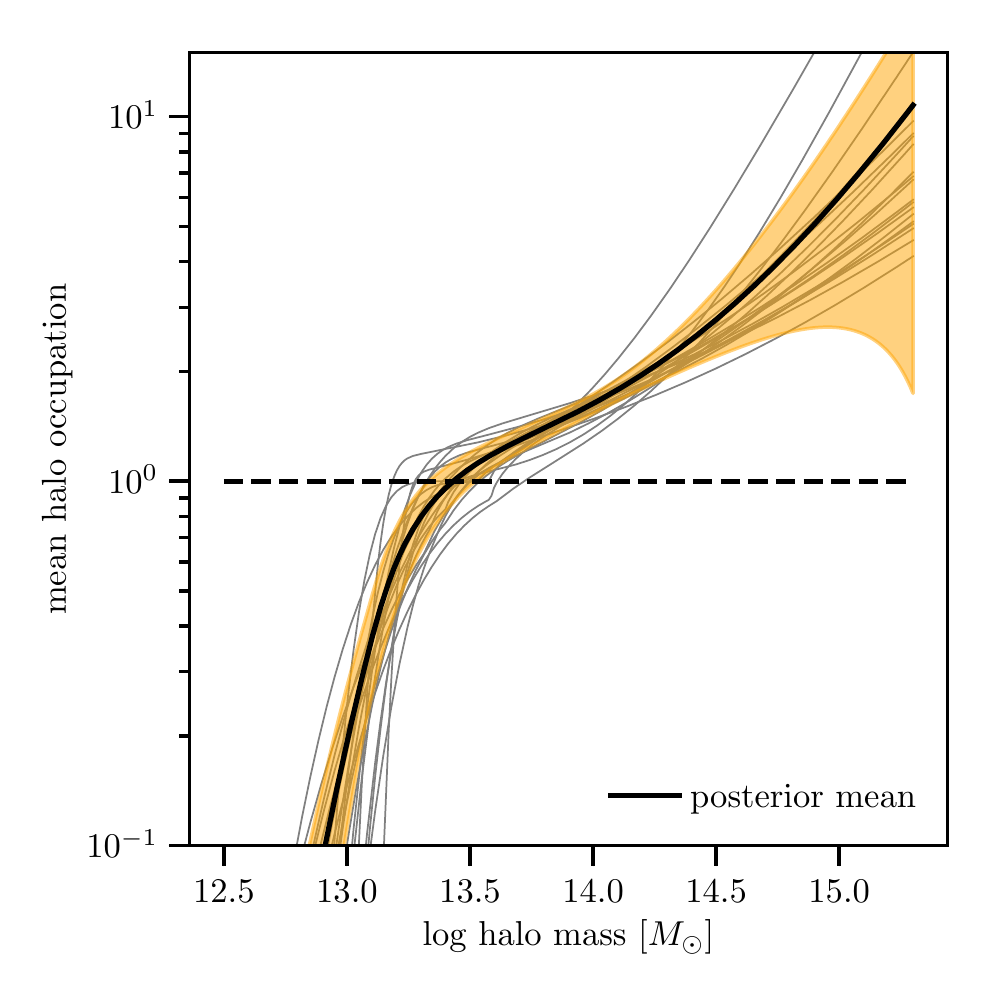}
 \caption{Posterior predictive distribution for $\langle N | M \rangle$ when fit to LOWZ data. The solid black line is the posterior mean, with the standard deviation plotted as the pale orange band around the black line. Individual model HODs are drawn at random from the posterior samples and plotted as gray lines. The halo occupation is only contrained at the factor of 2 level (or worse) in the cluster mass regime.}
  \label{fig:ppd_hod_data}
\end{figure}

Figure \ref{fig:ppd_hod_data} shows the posterior predictive distribution of the mean halo occupation. The shape of this distribution is fairly well constrained, but the occupation in rare high mass halos is uncertain and many of the individual HOD parameters are poorly constrained (Table \ref{table:posterior_mocks}). These uncertainties may appear surprising when compared to the per cent-level constraints on some halo occupation parameters reported previously in the literature (e.g. \citealt{Zehavi_2011,Sinha_2018}), but it was in fact predicted by our forecasts in Paper I. Some of the additional uncertainty is undoubtedly due to marginalizing over cosmological parameters in addition to HOD parameters, which has been seldom done in previous analyses (and never, to our knowledge, in the context of making halo model predictions by populating $N$-body simulations). The uncertainties in individual parameters also depends on the choice of parameterization because of degeneracies, so a representation like Figure \ref{fig:ppd_hod_data} is more informative in terms of both the mean and uncertainties of the inferred halo occupation. A parameterization which more tightly follows the posterior predictive distribution of halo occupation as a function of mass (Figure \ref{fig:ppd_hod_data}) may be worth exploring in future work.

\begin{figure}
  \includegraphics[width=\columnwidth]{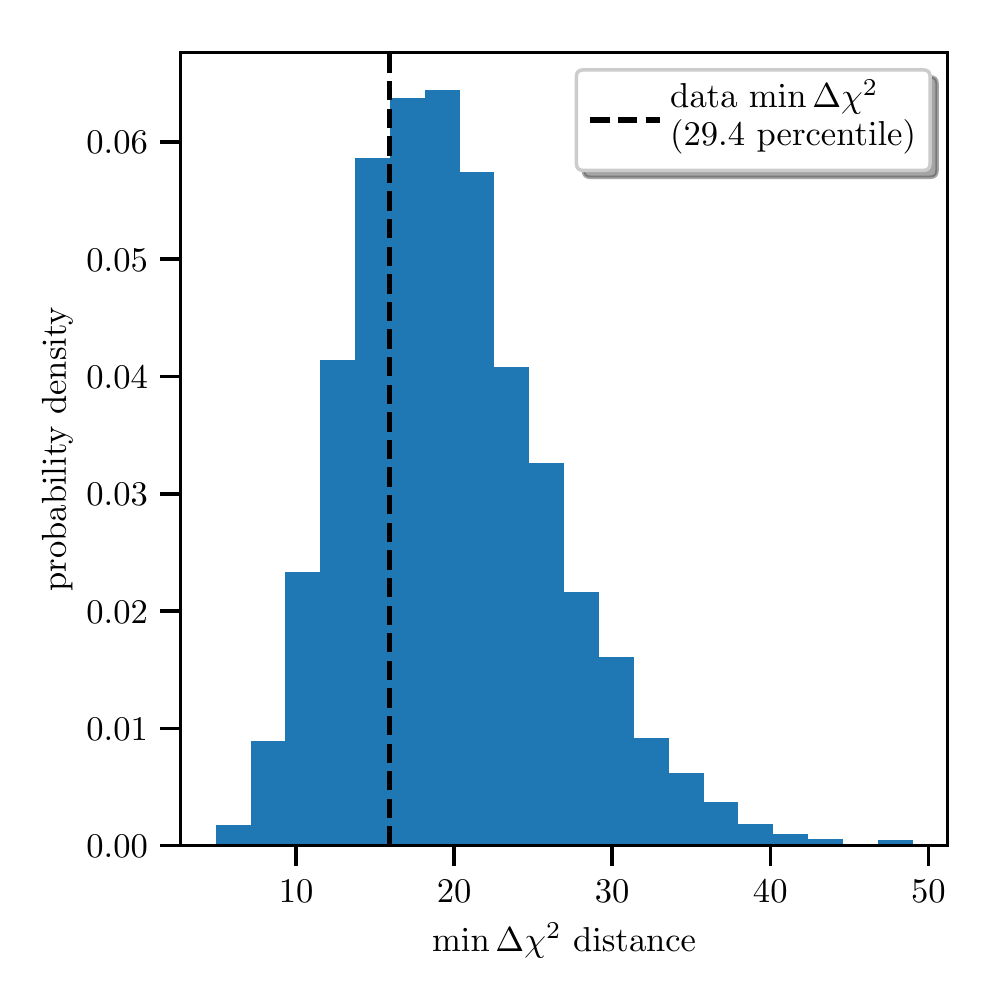}
\caption{Posterior discrepancy distribution for our emulator-based model when fit to LOWZ data. The value of $\Delta \chi^2$ computed for the data is shown as the vertical dashed black line. This value lies well within the distribution expected for data drawn from the posterior (blue histogram), and thus shows that there is no significant discrepancy between the model and data.}
  \label{fig:discrepancy_data}
\end{figure}

Finally, we compute the posterior discrepancy distribution for our fiducial analysis on data just as we did previously for the analyses on mock lightcones (Figure \ref{fig:discrepancy_data}), following the methodology advocated by \cite{Gelman_1996}. The $\text{min} \, \Delta\chi^2$ discrepancy computed for the data lies near the peak of the expected distribution, computed assuming that the data is drawn from the posterior samples. As before, we compute the minimum $\Delta\chi^2$ between the data and the posterior samples to obtain $\text{min} \, \Delta\chi^2$ of the data. For the expected distribution of $\text{min} \, \Delta\chi^2$, we choose posterior samples at random and add noise according to our fiducial covariance matrix, and then compute the minimum $\Delta\chi^2$ between this synthetic observation and the set of posterior samples.  Interpreted as a frequentist statistical test, the value of $\text{min} \, \Delta\chi^2$ of our data is such that we do not reject the null hypothesis that our data is drawn from the posterior distribution.  While this test does not imply anything about the correctness of our model, it does suggest that our model is an adequate description of the data given its statistical precision, a desirable and nontrivial property of any forward modeling framework.

\section{Discussion}
\label{section:conclusions}

\subsection{Comparison with previous work}
\label{section:comparison}

Our posterior constraints on $S_8$ are in good agreement with those of \cite{Singh_2018}, who use the same data but using a minimum scale of $1 \, \hMpc$ and with a parameterized model of the cross-correlation coefficient between galaxies and matter inspired by that obtained in simulations instead of a full forward model of the galaxy clustering and galaxy-galaxy lensing, as used in this work.  While they conduct multiple analyses with various estimators and scale cuts, their statistical precision is comparable to ours, as they obtain (for their tightest-constrained value) $S_8 = 0.823 \pm 0.035$, a $4.2$ per cent uncertainty.  They do not include systematic uncertainties directly in deriving posterior parameter constraints but rather quote an additional systematic uncertainty of 6 per cent in addition to their quoted statistical uncertainty.  As a result, it is difficult to directly compare whether our analysis results in more precise constraints, but we note that when excluding systematic uncertainties from our analysis (i.e. removing the point mass term and fixing $A_{\text{lensing}} = 1$) we obtain a precision of 2.8 per cent on our posterior value of $S_8$, suggesting that including sub-Mpc scales and explicitly modeling both the clustering and lensing signal may provide additional cosmological constraining power in our analysis.

Our results may be expected to be very similar, as our halo occupation models predict similar values of the scale-dependent galaxy-matter cross-correlation coefficient assumed in the parameterization of \cite{Singh_2018}, given that the galaxy-galaxy lensing signal can be exactly decomposed into a product of the projected galaxy-matter cross-correlation and the projected galaxy bias \citep{Baldauf_2010}.  Galaxy assembly bias within the subhalo abundance matching framework does not significantly alter (at the percent level) the scale-dependent galaxy-matter cross-correlation coefficient \citep{McEwen_2016} and thus both approaches are relatively robust to the possible presence of galaxy assembly bias.

Our results also exhibit essentially no tension with the combined galaxy clustering, galaxy-galaxy lensing and cosmic shear analysis of \cite{DESY1KP}, as readily seen in Figure \ref{fig:posterior_data}. This analysis includes all $w$CDM parameters and additionally marginalizes over the effective number of neutrino species $N_{\text{eff}}$ as well of the sum of neutrino masses $\sum m_{\nu}$, whereas we fix $N_{\text{eff}}$ to the standard model value (3.046) and assume massless neutrinos. We also compare with an alternative analysis performed by \cite{DESY1KP} (section VIID) wherein they assume $\Lambda$CDM and fix the sum of neutrino masses $\sum m_{\nu}$ to the minimal value allowed by neutrino oscillation experiments (0.06 eV). We have explicitly verified from their posterior samples that these priors leave the $S_8$ (as defined in this work) constraint essentially unchanged compared to their fiducial $w$CDM priors, which marginalize over neutrino mass, and our results are likewise fully compatible with their alternative analysis with fixed neutrino mass.

Our results are also qualitatively similar to those of \cite{Leauthaud_2017}, who in addition to conducting a qualitative investigation of the galaxy-galaxy lensing signal in Planck-normalized galaxy mocks of the BOSS CMASS sample (with effective redshift $z_{\text{eff}} \approx 0.57$), perform a fitting-function-based \citep{vdBosch_2013} HOD analysis (without including possible central galaxy miscentering or incompleteness) to obtain a constraint on $S_8$ that is lower than that of the fiducial analysis of \emph{Planck} data at $2-3 \, \sigma$ significance (see their Figure 9 and accompanying text). Our HOD model is fully self-consistent and does not require any additional nuisance parameters to describe the `halo exclusion' effect, unlike their analysis, which may contribute to the comparatively greater statistical significance of the tension of our value of $S_8$ with that of \emph{Planck}. Our result also uses the BOSS LOWZ galaxy sample instead of the BOSS CMASS galaxy sample, which may be more robust to galaxy selection effects altering the form of the halo occupation, since the stellar mass completeness of LOWZ is greater over a wider redshift range than that of CMASS \citep{Leauthaud_2016}, although a definitive statement would require convincing modeling of the complicated redshift- and color-dependent selection effects.

\begin{figure}
  \includegraphics[width=\columnwidth]{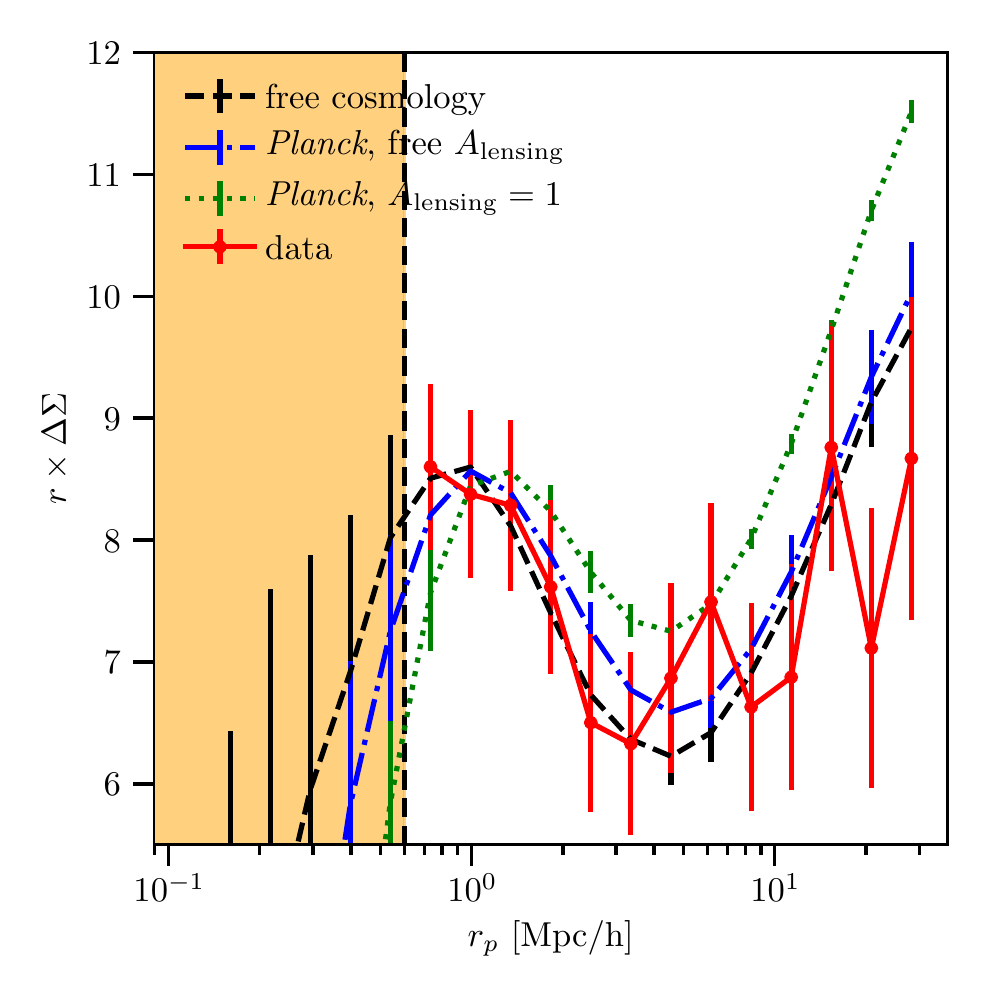}
\caption{Posterior predictive distribution for our emulator-based model when fit to LOWZ data. The data is shown as the red solid line with circles plotted for individual data points. The fiducial model is shown as the dashed black line. The model assuming the \emph{Planck} cosmology is shown as the dotted-dashed blue line, while the model assuming the \emph{Planck} cosmology and also fixing $A_{\text{lensing}}=1$ is shown as the dotted green line.}
  \label{fig:ds_comparison}
\end{figure}

\begin{figure}
	\includegraphics[width=\columnwidth]{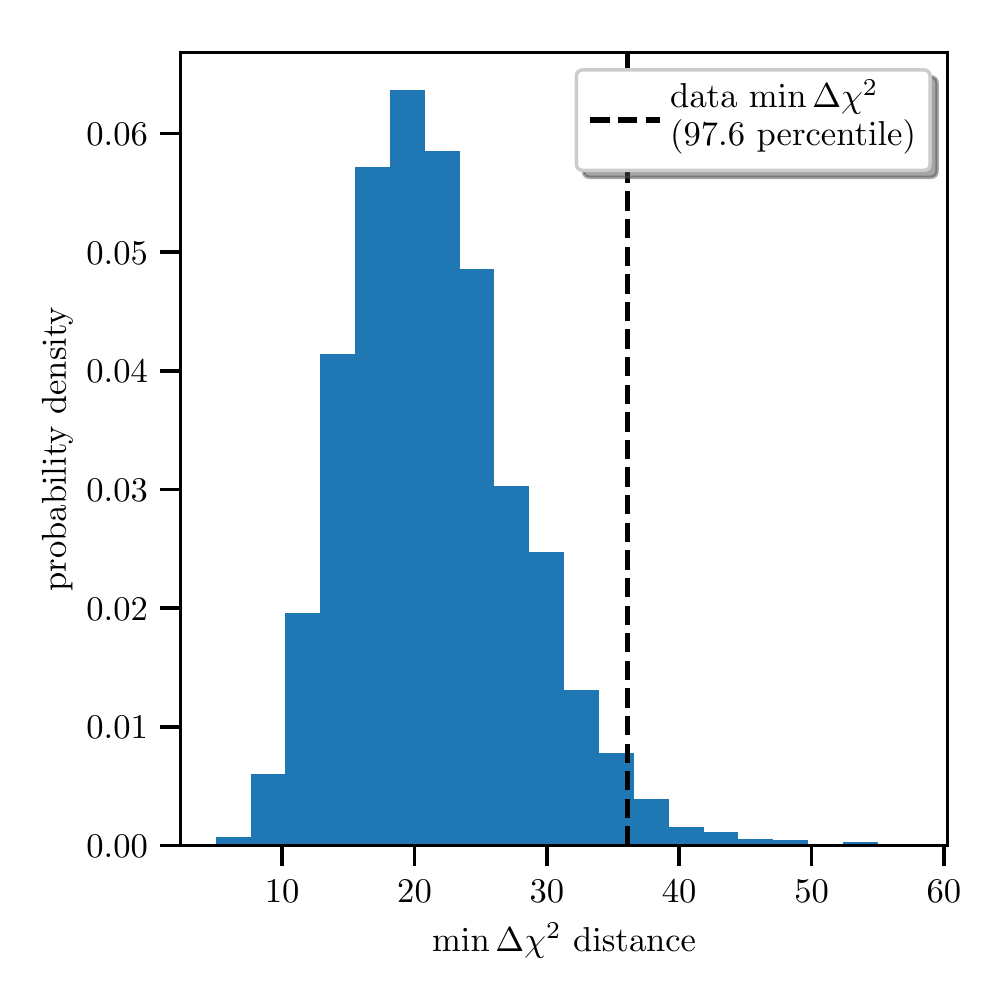}
\caption{Posterior discrepancy distribution for the \emph{Planck}-normalized cosmological model fit to the LOWZ data. We use the same discrepancy measure as in Figure \ref{fig:discrepancy_data} but instead computed for a model with the cosmological parameters fixed to the fiducial $\Lambda$CDM Planck Collaboration (2018) values. Interpreted as a single-tailed statistical test, this discrepancy measure suggests that the posterior of this model is in $\sim 2 \sigma$ tension with the data.}
\label{fig:discrepancy_planck}
\end{figure}

In Figure \ref{fig:ds_comparison}, we show the posterior predictive distribution for $\Delta\Sigma$ for three separate analysis cases.  The first (dashed black line) is the posterior mean derived from our fiducial analysis, with a free point mass parameter and a lensing amplitude parameter $A_{\text{lensing}}$ with a 6 per cent prior uncertainty derived from our systematic error estimates.  The second (dashed-dotted blue line) is the posterior mean derived from assuming that the cosmological parameters are fixed to the \emph{Planck} best-fit $\Lambda$CDM values, but still allowing the lensing amplitude to vary subject to the same prior as before.  The third (dotted green line) is the posterior mean derived from assuming that the \emph{Planck} cosmological parameter values are correct and fixing the lensing amplitude parameter $A_{\text{lensing}} = 1$. The first two analyses are in qualitative agreement with the data (red line), but the third analysis is not.  This comparison illustrates that the most significant (in terms of fractional amplitude) discrepancy between the data and the model is at large scales when the cosmological parameters are forced to agree with the \emph{Planck} results. We hypothesize that the contrary conclusion drawn by \cite{Lange_2019} may be due to their lack of point mass term in modeling the lensing signal.

In Figure \ref{fig:discrepancy_planck}, we present an alternative method of characterizing the tension between our results and the fiducial \emph{Planck} cosmology. We use the discrepancy distance measure described in section \ref{section:mocks} to compute the expected distribution of $\Delta \chi^2$ assuming that the data is drawn from the posterior of a \emph{Planck}-normalized model. The $\Delta \chi^2$ computed for the data lies at the 97th percentile of the expected distribution. Interpreted as a $p$-value, this corresponds to $p = 0.024$ or a $\sim 2\sigma$ tension when interpreted as a single-tailed statistical test. This measure of tension between the data and the \emph{Planck} cosmology may be more conservative than the standard comparison of the marginalized posterior of $S_8$ but it may be more robust to the (somewhat arbitrary) choice of priors on individual parameters and related effects due to the prior volume scaling with the number of nuisance parameters used in the analysis.

\subsection{Observational systematics}
\label{section:systematics}
Since we restrict our analysis to scales $\lesssim 30 \, \hMpc$, the dominant systematic uncertainty for $w_p$ is fiber collisions, which we avoid the need to explicitly model by adopting the standard fiber collision weights given in the LOWZ catalogs \citep{Ross_2014} and only using projected scales greater than twice the fiber collision scale at the maximum redshift of the LOWZ sample ($\gtrsim 0.6 \, \hMpc$).  This is shown to be sufficient for $\lesssim 1$ per cent accuracy of the projected correlation function in tests against mocks with synthetic fiber collisions by \cite{Guo_2012} and \cite{Yang_2019}. We neglect relativistic effects on clustering measurements, which are likely negligible for current datasets.

As discussed in \cite{Singh_2018}, the dominant source of uncertainty for lensing measurements used here is photo-z calibration ($\sim 5$ per cent), followed by shear calibration ($\sim 2$ per cent), intrinsic alignments (consistent with zero at $\sim 1-2$ per cent precision, according to the analysis of \citealt{Blazek_2012} on the same dataset used here), and magnification bias (estimated by \citealt{Singh_2019} to be $\lesssim 1$ per cent). We have not attempted to estimate reduced shear corrections or any additional higher-order lensing effects.  We note that fourth-order and higher lensing effects have been shown to be subdominant to shape noise for cosmic shear measurements even at LSST precision \citep{Petri_2017}.  We have used the multiplicative calibration parameter $A_{\text{lensing}}$ on the amplitude of $\Delta\Sigma$ in order to account for the combined effects of photo-z calibration bias, shear calibration bias, intrinsic alignments, and magnification bias, with a Gaussian prior centered at unity with a 6 per cent standard deviation.

\subsection{Baryonic and neutrino effects}
\label{section:baryons}

\cite{Singh_2018} conducted extensive parameter recovery tests by fitting to hydrodynamic simulations with strong AGN feedback and found that baryonic effects only affected $S_8$ by at most $\sim 2-3$ per cent, with their fits using a minimum scale of $1 \, \hMpc$.  However, when including shape noise in their analysis covariance matrix, they find essentially no bias in recovered $S_8$. We conducted identical tests, but down to our minimum scale of $0.6 \, \hMpc$, by modifying the `observed' $\Delta\Sigma$ from the fiducial mock lightcone by the ratio of $\Delta\Sigma$ (measured with $\Pi_{\text{min}}=0.01 \, \hMpc$; eq. \ref{eq:deltasigma}) with and without baryonic effects as measured in the original Illustris simulation by \cite{Singh_2018}.  We find that when the point mass is allowed to be negative, the effect of baryons on $S_8$ is consistent with zero ($\lesssim 0.6$ per cent difference from the true value of $S_8$ in our lightcones). Red contours in Figure \ref{fig:posterior_mocks} show the parameter constraints recovered from the mock catalog with this treatment of baryonic effects.

For future surveys with increased statistical precision, the best way to test for baryonic effects is likely to be cross-correlations with probes of hot gas, such as thermal Sunyaev-Zeldovich (tSZ) maps from Planck and tSZ and kinetic Sunyaev-Zeldovich (kSZ) maps from Stage IV cosmic microwave background experiments (CMB-S4; \citealt{Abazajian_2016}). With additional model flexibility, combining this cross-correlation with the cosmic shear correlation functions should enable self-calibration of baryonic effects on the matter distribution, or otherwise provide a convincing null test of their absence.

Clustering effects of massive neutrinos are not included in our emulator, as it has only recently become feasible to routinely simulate the effects of massive neutrinos on nonlinear matter clustering thanks to algorithmic advances \citep{Bird_2018,Banerjee_2018}.  The tests of \cite{Leauthaud_2017} at slightly higher redshift ($z \sim 0.57$) suggest that massive neutrinos may affect the matter clustering at the $5-10$ per cent level on the scales used in this work, with an effect size depending on the (currently unknown) value of the sum of the neutrino masses.  The best current limits are $\lesssim 0.1$ eV, assuming $\Lambda$CDM cosmology, but these relax to $\lesssim 0.3$ eV within $w$CDM cosmology \citep{Alam_2017,Choudhury_2018}. The effect of neutrino mass is strongly, but not exactly, degenerate with $S_8$ \citep{Ichiki_2012}.  Future emulators would be well advised to include such effects, especially given the strong discrepancy we infer between the low-redshift amplitude of matter clustering and that of Planck CMB measurements in the absence of massive neutrino effects.

\subsection{Halo model uncertainties}

Central galaxies may not be at the centers of their halos (e.g., \citealt{Ho_2009}), and thus the lensing signal may be mismatched with the clustering signal for this reason.  A model for miscentering and galaxy sample incompleteness was included in a fitting-function-based HOD cosmological analysis of CMASS galaxy clustering and galaxy-galaxy lensing by \cite{More_2015}, with posterior constraints indicating a $\sim 2 \sigma$ detection of a miscentered population of central galaxies. Since the LOWZ sample also consists of luminous red galaxies, it may be worthwhile to parameterize the miscentering effects in LOWZ galaxies in future work (but see \citealt{Lange_2019} for a counterargument regarding the plausibility of such effects).

In Paper I we introduced a parameter $Q_{\text{env}}$ to represent the possible effects of galaxy assembly bias (e.g. \citealt{Zentner_2016}). We found that $Q_{\text{env}}$ was not needed to obtain unbiased cosmological results from the SHAM lightcone (consistent with the result of \citealt{McEwen_2016}), and we therefore did not include it in our analysis.  The real universe may exhibit galaxy assembly bias in a form unlike that of SHAM, so it may be desirable to include an explicit parameterization of assembly bias in future cosmological analyses. Several alternate forms of assembly bias were explored by \cite{Yuan_2019}, who found that none of the candidate forms of assembly bias could explain the lensing amplitude of BOSS CMASS galaxies in a {$\Lambda$}CDM cosmology with \emph{Planck}-compatible parameters. Taken together, these results disfavor assembly bias as a solution to the observed lensing tensions.

\subsection{Possible empirical tests of robustness}

Confidence in small scale measurements of cosmological parameters as described here will require either Bayesian model averaging over a wide range of phenomenological models of galaxy bias (i.e., for a set of models which are considered equally likely \emph{a priori}, weighting the posteriors for the parameters in common between all models under consideration by the evidence integral of each model; see \citealt{Marshall_2003,Liddle_2006,Parkinson_2010,Vardanyan_2011} for applications to cosmological parameters) or will require the development of null tests on data, not just simulations.\footnote{As justification, we invoke Cromwell's dictum (as named by Bayesian statisticians): ``I beseech you, in the bowels of Christ, consider it possible that you are mistaken'' (quoted in \citealt{Rasmussen_2006}).}  While it is not possible to compute all models, we hope that our study will encourage model comparison and model averaging to be attempted with a finite number of models on LOWZ galaxy-galaxy lensing.\footnote{However, we note that any such comparison is subjective, since the evidence of a given model depends sensitively on the prior adopted on the parameters of that model, much more so than the posterior distribution itself (e.g., \citealt{Gelman_2017}). There exist non-Bayesian model averaging methods which are more robust to prior choices, e.g. using hyperparameter weights \citep{Lahav_2000,Trotta_2008}.}

When one is limited to evaluating a single model, or when it is unknown whether the correct model lies within the set of models under consideration, a reasonable test of model robustness is to fit multiple galaxy samples, showing one can recover the same cosmological parameters \emph{and} also fit the cross-correlation functions between the galaxy samples without significant discrepancies between model and data. This would likely require luminosity- and color-dependent modeling of the galaxy population.  (In principle, one would also need to include subsamples split on all of the galaxy properties used for sample selection.) This is an ambitious program of research, which will no doubt require the work of many people.\footnote{For usage of this phrase in an entirely different context, see \cite{Teller_1955}.}

\section{Summary and Outlook}

We have extended the methodology of \cite{Wibking_2019} (Paper I)
to enable emulation of galaxy clustering and GGL for $w$CDM cosmologies, sampling the cosmological parameter space allowed by \emph{WMAP} and \emph{Planck} CMB measurements and HOD parameters characteristic of the BOSS LOWZ sample of massive galaxies at $z=0.16-0.36$. We use the \cite{Garrison_2017} suite of 40 $w$CDM $N$-body simulations with matched Fourier phases, supplemented by 20 simulations of a fiducial \lcdm\ cosmology with varying phases which we use for covariance matrix calculations and for a sample variance correction to the mean model predictions.  We construct a Gaussian process emulator for the GGL observable $\Delta\Sigma(r_p)$ and for the ratio of the projected galaxy correlation function $w_p(r_p)$ to the prediction of an analytic halo model calculation. This ratio changes much more slowly with parameters than $w_p(r_p)$ itself, enabling us to achieve percent-level emulator accuracy over a wider range of parameters. We train the emulator by maximizing the leave-one-out pseudo-likelihood (Appendix \ref{appendix:GP}).

We test our approach on the light-cone mock catalogue of the BOSS LOWZ sample constructed by
\cite{Singh_2018}. This catalogue uses subhalo abundance matching rather than an HOD prescription to populate the simulated dark matter distribution with galaxies, so it tests the ability of an HOD-based emulator method to derive unbiased results from a catalogue that incorporates
a different model of nonlinear galaxy bias as well as evolution over the redshift range of the LOWZ sample.  When fitting the mock catalogue we adjust the covariance matrix in a way that 
effectively marginalizes over a central point mass \citep{Maccrann_2019},
i.e., an additive $1/r_p^2$ contribution to $\Delta\Sigma(r_p)$ with arbitrary normalization.  This parameter accounts for finite resolution in our N-body simulations, and it can also account for baryonic physics effects and for the possibility that central galaxies do not lie at the potential minima of their parent halos. Our HOD parameterization also includes a novel parameter $R_{\text{rescale}}$ that multiplies all halo virial radii by a constant factor; this parameter reduces sensitivity to the halo definition.

With fiducial parameter choices in the mock catalogue, we recover the true input value of the cosmological parameter combination $S_8 \equiv (\sigma_8/0.8228)(\Omega_m/0.3107)^{0.6}$ to within 1.7 per cent, about half of the estimated $1\sigma$ statistical uncertainty.  We constructed alternative mock catalogues with higher or lower satellite fractions, central galaxy incompleteness,
or baryonic effects on $\Delta\Sigma(r_p)$, and we again found unbiased recovery of $S_8$.  Because we have only a single mock LOWZ volume, we cannot average over many realizations to test for biases that are small compared to the statistical error.

After completing all mock catalogue tests and finalizing modeling choices,
we applied our method to the 
\cite{Singh_2018} measurements of LOWZ $w_p$ and $\Delta\Sigma$.
We find $S_8=0.847\pm 0.037$, a $3.5\sigma$ tension with the
Planck-normalized value of $1.00 \pm 0.02$ for a \lcdm\ cosmology.
A significant fraction of the 4.4 per cent error budget comes from systematic
uncertainty in the weak lensing measurements, modeled by a multiplicative
parameter $A_{\text{lensing}}$ with a 6 per cent Gaussian prior to represent estimated
uncertainties in photometric redshift zero points and shear calibration.
With $A_{\text{lensing}}$ fixed to one, our posterior weighted estimate would be
$S_8 = 0.831 \pm 0.029$, a $3.5$ per cent statistical error.
The point mass marginalization also contributes signficantly to the error budget.  If we fixed both $A_{\text{lensing}}$ and the point mass parameter, our statistical error on
$S_8$ would improve to 2.8 per cent.  If we instead force $S_8=1$ and
adopt our fiducial prior on $A_{\text{lensing}}$, we can fit the clustering and GGL
data with $A_{\text{lensing}} \approx 0.85$, implying that reproducing the \emph{Planck} $S_8$
normalization with our modeling choices requires a lensing systematic error that is about 2.5 times higher than estimated by \cite{Singh_2018}.

Our results are in excellent agreement with those of 
\cite{Singh_2018}, who model the same data with a different method, adopting
a parameterized galaxy-matter cross-correlation coefficient constrained by their mock catalogues. They are in qualitative agreement with the findings of
\cite{Leauthaud_2017}, who analyzed the higher redshift BOSS CMASS
sample using deeper weak lensing data over much smaller area.
Our normalization of $S_8$ is consistent with (but lower than)
that derived from the DES Y1 $3\times 2$pt analysis \citep{DESY1KP}.
Our measurement joins others (e.g., \citealt{Leauthaud_2017}; \citealt{Singh_2018}; \citealt{Mandelbaum_2013}; \citealt{Hildebrandt_2017}) that suggest an amplitude of low redshift matter clustering significantly below that of a \emph{Planck}-normalized \lcdm\ model.

In future work, several aspects of our methodology could be improved:
\begin{itemize}
\item the calculation of the redshift-space distortion correction to $w_p$ could be made more efficient and properly cosmology-dependent via the use of the FFTLOG algorithm \citep{Hamilton_2000} to compute the integrals required in the \cite{Kaiser_1987} model;
\item the sample variance of the emulator could be further reduced via $N$-body simulations initialized with `fixed and paired' phases, as advocated by \cite{Angulo_2016};
\item for a fixed number of samples, the emulator accuracy might be improved with a sampling design that incorporates both clustered and volume-filling subsamples of the parameter space \citep{Zhu_2006,Zimmerman_2006};
\item future datasets may be precise enough as to require explicit redshift-dependent modeling, which may be straightforwardly incorporated via an additional label dimension of the training data;
\item halo definition may be marginalized over in a fully self-consistent way by introducing the spherical overdensity threshold as an emulator parameter and sampling from this parameter space jointly with the cosmological parameters;
\item the robustness of the cosmological results may be empirically tested by analyzing disjoint subsamples split by galaxy properties used in target selection, such as luminosity and color;
\item the effect of massive neutrinos on nonlinear matter clustering may be significant at the $\sim 5$ per cent level in $\Delta\Sigma$ \citep{Leauthaud_2017} and should be included in $N$-body simulations with the benefit of recently-developed numerical techniques \citep{Bird_2018,Banerjee_2018}; and
\item although fourth-order and higher effects are likely negligible for any planned weak lensing experiment (\citealt{Petri_2017}), weak lensing effects arising at third order in the gravitational potential such as reduced shear and magnification bias corrections (e.g. \citealt{Krause_2010}) may need to be explicitly modeled in future analyses via emulation of the three-point function.
\end{itemize}

In the near future, expanded data sets from DES, KiDS, and HSC
will allow this lensing amplitude tension to be tested at higher precision using
both cosmic shear and galaxy clustering + GGL.
The principal challenge will be controlling weak lensing systematics
at the necessary level, though further tests of the modeling methods
will also be needed as the statistical precision improves.
GGL and cosmic shear have different quantitative response to
biases in photometric redshifts, shear calibration, and galaxy
intrinsic alignments, so they provide useful cross-checks as
well as increased statistical precision in combination.  
GGL + clustering analyses of galaxy samples with distinct clustering
properties and at multiple redshifts allow further tests
for the robustness of cosmological parameter constraints.
If future observations indeed show the need for physics beyond
that of GR+\lcdm, then GGL+clustering analysis will be a powerful 
tool for measuring its redshift, scale, and environment dependence.

\section*{Acknowledgements}

We thank Niall MacCrann and the DES Collaboration for providing the posterior samples for the DES `fixed-neutrino' analysis used in section \ref{section:comparison}.

BDW thanks Chris Hirata for helpful
discussions of covariance matrices and many other topics,
Joe McEwen and Xiao Fang for explaining the \textsc{FFTLOG} algorithm and
its application to one-loop standard perturbation theory, 
Martin White for suggesting empirical tests of the robustness of small-scale
measurements of cosmological parameters, and Andreas Berlind for first
suggesting to him the possibility of using small-scale galaxy clustering
to measure cosmological parameters.

BDW is supported by the
National Science Foundation Graduate Research Fellowship Program under
Grant No. DGE-1343012. ANS is supported by the Department of Energy
Computational Science Graduate Fellowship Program of the Office of
Science and National Nuclear Security Administration in the Department
of Energy under contract DE-FG02-97ER25308.  BDW, ANS, and DHW are supported
in part by NSF grant AST-1516997. Any opinions,
findings, and conclusions or recommendations expressed in this material are those of the
author(s) and do not necessarily reflect the views of the National
Science Foundation. DJE is supported by U.S. Department of Energy grant
DE-SC0013718 and as a Simons Foundation Investigator.  LG is supported by
the Simons Foundation.

Simulations were analyzed in part on computational resources of the
Ohio Supercomputer Center \citep{OhioSupercomputerCenter1987},
with resources supported in part by the Center for Cosmology and AstroParticle
Physics at the Ohio State University.

\emph{Software:} \textsc{matplotlib} \citep{Hunter_2007}, GNU Scientific Library
\citep{GSL_2009}, Corrfunc \citep{Sinha_2017}.
BDW especially thanks the authors of the \textsc{Corrfunc} pair-counting code,
without which this project (and his Ph.D. thesis) would not have been feasible.

This research has made use of NASA's Astrophysics Data System.



\bibliographystyle{mnras}
\bibliography{bibliography/biblio}



\appendix

\section{Analytic halo model}
\label{appendix:halo_model}

Although we use a numerical halo model for our predictions, we need an analytic halo model
in order to reduce the dynamic range of the emulated quantity and increase our emulation accuracy.
For this purpose, we construct a simple analytic halo model that does not attempt to model
halo exclusion or residual RSD effects.  Although we only use the analytic halo model in order to
compute the ratio $w_{\text{p,sim}} / w_{\text{p,analytic}}$ in this work, we include the equations for $\Delta\Sigma$ for completeness. For notational clarity, we drop the `analytic' subscripts used elsewhere in this work in order to indicate the use of the halo model described in this Appendix.

We write the configuration-space correlation functions as a the sum of a 1-halo term and a 2-halo term
\begin{align}
\xi(r) = \xi_{\text{1h}}(r) + \xi_{\text{2h}}(r) \, ,
\end{align}
where the 1-halo term $\xi_{\text{1h}}$ arises from galaxy pairs (or galaxy-matter `pairs') within a given halo and the 2-halo term $\xi_{\text{2h}}$ arises from galaxy pairs (or galaxy-matter `pairs') between two distinct halos.

\subsection{1-halo term $\xi_{\text{1h}}$}

In writing the 1-halo term, we follow the real-space formulation of \cite{Zheng_2007}. This allows for faster computation of the 1-halo term than its Fourier-space formulations (e.g., \citealt{vdBosch_2013}).

\subsubsection{Galaxy autocorrelation}
The 1-halo galaxy autocorrelation is given by the normalized differential pair counts of galaxies within halos \citep{Berlind_2002}, assuming Poisson satellite counts:
\begin{align}
1 + \xi_{1h}(r) = \frac{{DD}_{\text{cs}}(r) + {DD}_{\text{ss}}(r)}{ RR(r) }
\end{align}
where
\begin{align}
{DD}_{\text{cs}}(r) = &\int_{0}^{\infty} \, \langle N_{\text{cen}}(M_h) \rangle \, \langle N_{\text{sat}}(M_h) | N_{\text{cen}} = 1 \rangle \nonumber\\
&\times I'\left(\frac{r}{R_{\text{vir}}(M_h)}, \, c_{\text{vir}}(M_h)\right) \, \frac{dn}{dM_h} \, \frac{1}{R_{\text{vir}}(M_h)} \, dM_h \\
{DD}_{\text{ss}}(r) = &\int_{0}^{\infty} \, \frac{1}{2} \langle N_{\text{cen}}(M_h) \rangle \, \langle N_{\text{sat}}(M_h) | N_{\text{cen}} = 1 \rangle^2 \nonumber\\
&\times F' \left(\frac{r}{R_{\text{vir}}(M_h)}, \, c_{\text{vir}}(M_h)\right) \, \frac{dn}{dM_h} \, \frac{1}{R_{\text{vir}}(M_h)} \, dM_h
\end{align}
and
\begin{align}
{RR}(r) = 2\pi r^2 n_g^2 \, .
\end{align}
For an NFW profile, the differential (w.r.t dimensionless radius $r/R_{\text{vir}}$) pair count functions $I'$ and $F'$ are \citep{Sheth_2001}:
\begin{align}
I'(x, \, c) = 
	\begin{cases}
		\frac{ 1 }{ \ln(1+c) - \frac{c}{1+c} } \frac{ c x^2 }{ x \, (1+x)^2 } & 0 \leq x \leq 1 \nonumber \\
		0 & x > 1
	\end{cases}
\end{align}
and
\begin{align}
F'(x, \, c) = &\frac{ c^3 x^2 }{\left[ \ln(1+c) - c/(1+c) \right]^2} \nonumber\\
	&\times
	\begin{cases}
		\frac{ -4(1+a) \, + \, 2as(1+2a) \, + \, a^2 s^2 }{ 2s^2 (1+a)^2 (2+s) } \nonumber\\
		\, \, + s^{-3} \ln \left( \frac{ (1+a-as)(1+s) }{ 1+a } \right) \nonumber\\
		\, \, + \frac{ \ln(1+s) }{ s(2+s)^2 }
				& 0 \leq s \leq 1 \\
		\frac{ 1 }{ s(2+s)^2 }  \ln \left( \frac{ 1+a }{ as+a-1 } \right) \nonumber\\
		\, \, + \frac{ sa^2 - 2a }{ 2s(1+a)^2 (2+s) }
				& 1 < s \leq 2 \nonumber\\
		0 & s > 2
	\end{cases}
\end{align}
where $s = xc$ and $a = 1/c$, and the galaxy number density is
\begin{align}
n_{g} = \int_{0}^{\infty} \left( \langle N_{\text{cen}}(M_h) \rangle + \langle N_{\text{sat}}(M_h)\rangle \right) \, \frac{dn}{dM_h} \, dM_h \, .
\end{align}
Note that here $\langle N_{\text{sat}} \rangle$ is the fully marginalized satellite halo occupation (i.e., not conditioned on having a central in a given halo).

The virial radius is defined as
\begin{align}
R_{\text{vir}}(M_h) = \left( \frac{3 M_h}{4\pi \Delta_{\text{SO}} \rho_m} \right)^{1/3}
\end{align}
where $\rho_m = \rho_{\text{crit}} \Omega_m$ ($\rho_{\text{crit}}$ is the critical density of the universe) and we choose $\Delta_{\text{SO}} = 200$ for consistency with the halo mass-concentration relation described below (although this choice is one only of convenience).

We use the halo mass-concentration relation of \cite{Correa_2015}
\begin{align}
c_{\text{vir}} = 10^{\alpha + \beta \log M_h \left[ 1 + \gamma (\log M_h)^2 \right]}
\end{align}
where
\begin{align}
\alpha &= 1.62774 - 0.2458 \, (1 + z) + 0.01716 \, (1 + z)^2 \, , \\
\beta &= 1.66079 + 0.00359 \, (1 + z) - 1.6901 \, (1 + z)^{0.00417} \, , \\
\gamma &= -0.02049 + 0.0253 \, (1 + z)^{-0.1044} \, ,
\end{align}
and the mass function is the fitting formula of \cite{Tinker_2008}:
\begin{align}
\frac{dn}{dM_h}(M_h) = f(\sigma) \frac{\rho_m}{M_h} \frac{d \ln \sigma^{-1}}{dM_h} \, ,
\end{align}
where
\begin{align}
\sigma^2(M_h) &= \frac{1}{2\pi^2} \int_{0}^{\infty} dk \, k^2 \, P_{\text{lin}}(k) \, \left[ W(k,R_{\text{vir}}) \right]^2 \, , \\
W(k,R) &= \frac{ 3 \left[ \sin(kr) - kr \cos(kr) \right] }{ (kr)^3 } \, , \\
\frac{d \ln \sigma^{-1}}{d M_h} &=  \frac{1}{2\pi^2} \int_{0}^{\infty} \, dk \, k^2 P_{\text{lin}}(k) \, \frac{ W(k,R_{\text{vir}}) }{ -\sigma^2(M_h) } \,  \frac{dW}{dM_h}(k,R_{\text{vir}}) \, , \\
\frac{dW}{dM_h}(k,R) &= k M_h^{-2/3} \left(\frac{3}{4\pi\rho_m}\right)^{1/3} \nonumber\\
&\times \left( \frac{\sin (kR)}{(kR)^2} + \frac{3 \cos (kR)}{(kR)^3} - \frac{3 \sin (kR)}{(kR)^4} \right) \, ,
\end{align}
and
\begin{align}
f(\sigma) &= A \left[ \left(\frac{\sigma}{b}\right)^{-a} + 1 \right] \, e^{-c/\sigma^2} \, , \\
A &= 0.186 \, (1.0+z)^{-0.14} \, , \\
a &= 1.47 \, (1.0+z)^{-0.06} \, , \\
b &= 2.57 \, (1.0+z)^{-\alpha} \, ,  \\
c &= 1.19 \, .
\end{align}

\subsubsection{Galaxy-mass cross-correlation}
For the 1-halo term, we have
\begin{align}
1 + \xi_{\text{gm,1h}} &= \frac{DD_{\text{cm}} + DD_{\text{sm}}}{RR_{\text{gm}}} \, , \\
DD_{\text{cm}} &= \int_{0}^{\infty}  M_h \, \langle N_{\text{cen}} (M_h) \rangle \nonumber\\
&\times I' \left( \frac{r}{R_{\text{vir}}}, \, c_{\text{vir}}(M_h) \right) \, \frac{dn}{dM_h} \, \frac{1}{R_{\text{vir}}(M_h)} \, dM_h \, , \\
DD_{\text{sm}} &= \int_{0}^{\infty}  M_h \, \langle N_{\text{sat}} (M_h) \rangle \nonumber\\
&\times F' \left( \frac{r}{R_{\text{vir}}}, \, c_{\text{vir}}(M_h), \, A_{\text{conc}} c_{\text{vir}}(M_h) \right) \nonumber\\
&\times \frac{dn}{dM_h} \, \frac{1}{R_{\text{vir}}(M_h)} \, dM_h \, ,  \\
RR_{\text{gm}} &= 2\pi r^2 n_g \rho_m \, .
\end{align}

We write the convolution of an NFW profile with another NFW profile of differing concentration $F'(x,c_1,c_2)$ as (Appendix A, \citealt{Zheng_2007}), with $s = 2x$:
\begin{align}
F'(x, \, c_1, \, c_2) = \, &\sqrt{A_{\star}(c_1) A_{\star}(c_2)} \, s \nonumber\\
&\times	\begin{cases}
		f_1 + f_2 + f_3 + f_4	& x \leq 0.5 \\
		f						& 0.5 < x \leq 1 \\
		0						& x > 1
	\end{cases}
\end{align}
where
\begin{align}
f_1 = \, & (c_2 + c_1 + c_1 c_2 s)^{-2} \, \ln\left[ (1+c_1 s)(1+c_2 s) \right] \nonumber\\
&+ \frac{c_1 s}{c_2 (c_2 + c_1 + c_1 c_2 s) (1 + c_1 s) } \\
f_2 = \, & (c_2 - c_1 + c_1 c_2 s)^{-2} \, \ln\left[ (1+c_1 s)(1+c_2-c_2 s) / (1+c_2) \right] \nonumber\\
&- \frac{c_1 (1-s)}{ c_2 (c_2 - c_1 + c_1 c_2 s) (1+c_1 s) (1+c_1) } \\
f_3 = \, & (c_2 - c_1 - c_1 c_2 s)^{-2} \, \ln\left[ (1+c_2 s)(1+c_1-c_1 s) / (1+c_2) \right] \nonumber\\
&+ \frac{c_1(1-s)}{ c_2 (c_2 - c_1 - c_1 c_2 s) (1 + c_1 - c_1 s) } \\
f_4 = \, & \frac{ -s }{ c_2 (1 + c_1) (1 + c_2) (1 + c_1 - c_1 s) } \\
f = \, & (c_2 + c_1 + c_1 c_2 s)^{-2} \, \ln\left[ \frac{ (1+c_1) (1+c_2) }{ (1 - c_1 + c_1 s)(1 - c_2 + c_2 s) } \right] \nonumber\\
&+ \frac{ s - 2 }{ (1+c_1) (1+c_2) (c_2 + c_1 + c_1 c_2 s) } \, ,
\end{align}
where $A_{\star}$ is approximately
\begin{align}
A_{\star}(c) \approx \, &A_0 \, c^{3+\alpha} \, \left[ (1 + (c/c_T)^{ (\beta - \alpha)/\mu } \right]^{\mu} \nonumber\\
&\times \left[ 1 + B_0 \, \sin \left( \omega (\log c - \phi) \right) \right] \, ,
\end{align}
where
\begin{align}
A_0 &= 2.4575 \, , \\
\alpha &= -3.099 \, , \\
\beta &= 0.617 \, , \\
c_T &= 1.651 \, , \\
\mu &= 4.706 \, , \\
B_0 &= 0.0336 \, , \\
\omega &= 2.684 \, , \text{ and} \\
\phi &= 0.4079 \, .
\end{align}

\subsection{2-halo term $\xi_{\text{2h}}$}

For the linear matter power spectrum $P_{\text{lin}}(k)$, we use the `no-wiggles' fitting formula of
\cite{Eisenstein_1998}, which is substantially cheaper to compute than using Boltzmann codes, which makes our posterior inferences faster.

We use the halo mass-bias relation of \cite{Tinker_2010}:
\begin{align}
b(M_h) &= 1 - A \frac{ \nu^a }{ \nu^a + \delta_c^a } + B \nu^b + C \nu^c \, , \\
A &= 1 + 0.24 \, y \, e^{-(4/y)^4} \, \, \\
B &= 0.183 \, \, \\
C &= 0.019 + 0.107 \, y + 0.19 \, e^{ -(4/y)^4 } \, \, \\
a &= 0.44 \, y - 0.88 \, \, \\
b &= 1.5 \, \, \\
c &= 2.4 \, , \\
y &= \log \Delta_{\text{SO}} \, , \text{ and} \\
\nu(M_h) &= \frac{ \delta_c }{ \sigma(R_{\text{vir}}(M_h)) } \, ,
\end{align}
where $\delta_c = 1.686$. We obtain the overall large-scale bias of the galaxy sample as
\begin{align}
b_g = n_{g}^{-1} \int_{0}^{\infty} \frac{dn}{dM_h} \langle N | M_h \rangle \, b(M_h) \, dM_h \, .
\end{align}

\subsubsection{Galaxy autocorrelation}
For the galaxy autocorrelation between distinct halos, we have
\begin{align}
\xi_{\text{gg,2h}}(r) = b_g^2 \, \int_{0}^{\infty} \frac{4\pi k^2 dk}{(2\pi)^3} P_{\text{lin}}(k) \, j_0(kr) \, dk \, .
\end{align}

\subsubsection{Galaxy-matter cross-correlation}
For the galaxy-matter cross-correlation between distinct halos, we have
\begin{align}
\xi_{\text{gm,2h}}(r) = b_g \, r_{\text{gm}} \, \int_{0}^{\infty} \frac{4\pi k^2 dk}{(2\pi)^3} P_{\text{lin}}(k) \, j_0(kr) \, dk
\end{align}
where we further assume that $r_{\text{gm}} = 1$ on all 2-halo scales.

\section{Analytic covariance matrices}
\label{appendix:covariance}

Rewriting equation A2 of \cite{Krause_2017} for the three-dimensional power spectrum, we have:
\begin{equation}
    \begin{split}
       &\text{Cov}(P^{ij}_{AB}(k_1), \, P^{kl}_{CD}(k_2)) = 
        \frac{(2\pi)^3 \, \delta(k_1 - k_2) }{ V_s \, (4\pi k_1^2) } \\
        &\times \left[ 
           ( P^{ik}_{AC}(k_1) + \delta_{ik} \delta_{AC} N^i_A )
        \, ( P^{jl}_{BD}(k_2) + \delta_{jl} \delta_{BD} N^j_B ) \right. \\
        &+ \, \left. ( P^{il}_{AD}(k_1) + \delta_{il} \delta_{AD} N^i_A )
        \, ( P^{jk}_{BC}(k_2) + \delta_{jk} \delta_{BC} N^j_B ) \right] \, ,
    \end{split}
\end{equation}
where $\delta_{ij}$ refers to the Kronecker delta function and $N^i_A$ is the appropriate noise term for probe $A$ in redshift bin $i$ (e.g., for the 3D galaxy density field, this is $1/n^i_g$, where $n^i_g$ is the number density of galaxies in redshift bin $i$).

Specializing to the case of a single redshift bin (i.e., $i=j=k=l$), we have
\begin{equation}
    \begin{split}
       &\text{Cov}(P_{AB}(k_1), \, P_{CD}(k_2)) = 
        \frac{(2\pi)^3 \, \delta(k_1 - k_2) }{V_s \, (4\pi k_1^2) } \\
        &\times \left[ 
           ( P_{AC}(k_1) + \delta_{AC} N_A )
        \, ( P_{BD}(k_2) + \delta_{BD} N_B ) \right. \\
        &+ \, \left. ( P_{AD}(k_1) + \delta_{AD} N_A )
        \, ( P_{BC}(k_2) + \delta_{BC} N_B ) \right]
    \end{split}
\end{equation}
where the sum of products of power spectra (in brackets) is equal to the four-point function $\langle A B C D \rangle$ when $A$, $B$, $C$, and $D$, are Gaussian fields (compare with Eq. 4 of \citealt{Cooray_2001}).

For projected two-point functions with pairwise line-of-sight weight functions $W_{AB}$ and $W_{CD}$, we have
\begin{equation}
    \begin{split}
       &\text{Cov}^{2D}(P_{AB}(k_1), \, P_{CD}(k_2)) = 
        \int_{-\infty}^{\infty} d\Pi \, W_{AB}(\Pi) \, W_{CD}(\Pi) \\
        &\times \frac{(2\pi)^3 \, \delta(k_1 - k_2)}{V_s \, (4\pi k_1^2) } \\
        &\times \left[ 
           ( P_{AC}(k_1; \, \Pi) + \delta_{AC} N_A )
        \, ( P_{BD}(k_2; \, \Pi) + \delta_{BD} N_B ) \right. \\
        &+ \, \left. ( P_{AD}(k_1; \, \Pi) + \delta_{AD} N_A )
        \, ( P_{BC}(k_2; \, \Pi) + \delta_{BC} N_B ) \right] \, ,
    \end{split}
\end{equation}
where $\Pi$ is the relative line-of-sight distance between the points of a pair used in computing the two point statistic of interest.

Ignoring the finite-size correlation function bin width and survey boundary effects, for scalar projected correlation functions $w_{AB}(r_p)$ and $w_{CD}(r_p)$, we have
\begin{align}
\label{eq:covariance}
    \text{Cov}(w_{AB}(r_i), \, &w_{CD}(r_j)) = \int \int \frac{ dk_1^3 }{ (2\pi)^3 } \, 
    \frac{ dk_2^3 }{ (2\pi)^3 }\, e^{i\bm{k_1}\cdot\bm{r_i}} \, e^{i\bm{k_2}\cdot\bm{r_j}} \nonumber\\
	&\times \text{Cov}^{2D}\left(P_{AB}(k_1), \, P_{CD}(k_2)\right) \nonumber\\
    = &\frac{1}{V_s} \left[ \int d\Pi \, W_{AB}(\Pi) \, W_{CD}(\Pi) \right] \nonumber\\
	&\int_0^{\infty} \frac{k \, dk}{2\pi} \, J_0(k r_i) \, J_0(k r_j) \nonumber\\
    &\times \left[ 
           ( P_{AC}(k) + \delta_{AC} N_A )
        \, ( P_{BD}(k) + \delta_{BD} N_B ) \right. \nonumber\\
        &+ \, \left . ( P_{AD}(k) + \delta_{AD} N_A )
        \, ( P_{BC}(k) + \delta_{BC} N_B ) \right] \, ,
\end{align}
where $i,j$ here refer to the bin indices $r_{p,i}$ of the projected correlation functions.  When applying this expression to compute a covariance matrix, we stress that it is necessary to average the integrand over the bin width \emph{before} performing the outer integrals.  Otherwise, integrals of this form diverge whenever $N_A$ or $N_B$ are nonzero, due to the identity \citep{Jackson_1975,Gradshteyn_2007}:
\begin{align}
\int_0^{\infty} k \, J_{\nu}(ak) \, J_{\nu}(bk) \, dk = \frac{1}{a} \delta_D(b - a) \, .
\end{align}
Physically, this divergence represents the variance tending toward infinity as the number of galaxy pairs contained within an infinitesimal radial bin goes to zero.  The required bin-averaging of the integrand can be carried out by replacing each Bessel function $J_{\nu}(kr_i)$ with the bin-averaged Bessel function $\bar J_{\nu}(kr_i)$, where the average is taken over an annulus of inner radius $r_i$ and outer radius $r_{i+1}$.

\subsection{Covariance of $w_{p,gg}$}

For the projected two point function $w_{p,gg}(r_p)$, the pairwise line of sight weight function is
\begin{equation}
    W_{\text{gg}}(\Pi) = 
     \begin{cases}
        1 & |\Pi| \leq \Pi_{\text{max}} \\
        0 & \text{otherwise} \,
    \end{cases} 
\end{equation}
(compare with equation A21 of \citealt{Singh_2016}) and the integral over the weight function is
\begin{equation}
    \int_{-\infty}^{\infty} d\Pi \, W_{\text{gg}}(\Pi) \, W_{\text{gg}}(\Pi) = 2\Pi_{\text{max}} \, .
\end{equation}
Therefore eq. \ref{eq:covariance} becomes
\begin{equation}
\begin{split}
    \text{Cov}(&w_p(r_i), \, w_p(r_j)) = \frac{2}{V_s} \left[ \int d\Pi \, W_{\text{gg}}^2(\Pi) \right] \\
		&\times \int_0^{\infty} \frac{k \, dk}{2\pi} \, J_0(k r_i) \, J_0(k r_j) \left( \Pgg(k) + N_g \right)^2 \\
        = \, &\frac{4 \Pi_{\text{max}}}{V_s} \int_0^{\infty} \frac{k \, dk}{2\pi} \, J_0(k r_i) \, J_0(k r_j) \left( \Pgg(k) + \frac{1}{n_g} \right)^2 \, .
\end{split}
\end{equation}

\subsection{Covariance of $\gamma_t$}

For galaxy-galaxy lensing, after modifying the Fourier transform to account for the fact that the $\gamma_t$ is a component of a spin-2 tensor, eq. \ref{eq:covariance} reduces to 
\begin{align}
 \text{Cov}(&\gamma_t(r_i), \gamma_t(r_j)) = \frac{1}{V_s} \int d\Pi \, W_{g \gamma}^2(\Pi) \, \int_0^{\infty} \frac{k \, dk}{2\pi} \, J_2(k r_i) J_2(k r_j) \nonumber\\
	&\times \left[ 
           \left( P_{gg}(k) + N_g \right)
        \, \left( P_{\gamma \gamma}(k; \, \Pi) + N_{\gamma} \right)
        + P_{g \gamma}^2(k) \right] \nonumber\\
     = \, &\frac{1}{V_s} \int_0^{\infty} \frac{k \, dk}{2\pi} \, J_2(k r_i) J_2(k r_j) \nonumber\\
	 &\times \left[ 
           \left( \Pgg(k) + \frac{1}{n_g} \right)
        \, \left( \, P_{\gamma \gamma}^{2D}(k) + \frac{\sigma_{\gamma}^2}{\Sigma_s} \right)
        +  \Pi_{\text{lens}} P_{g \gamma}^2(k) \right]
\end{align}
where
\begin{equation}
    P_{\gamma \gamma}^{2D} = \int_0^{\chi_s} d\chi \, \left( \frac{\bar \rho}{\Sigma_c(\chi, \chi_s)} \right)^2 \, \Pmm\left(k \, \frac{\chi_l}{\chi}\right) \, ,
\end{equation}
\begin{equation}
    P_{g \gamma} \approx \left( \frac{\bar\rho}{\Sigma_c(\chi_l, \chi_s)} \right) \, \Pgm(k) \, ,
\end{equation}
and $\Pi_{\text{lens}}$ is the effective line-of-sight depth of the squared lensing weight function, given by
\begin{equation}
    \Pi_{\text{lens}} = \int_{0}^{\chi_s} d\chi \, \left( \frac{\Sigma_c(\chi_l, \chi_s)}{\Sigma_c(\chi, \chi_s)} \right)^2 \, .
\end{equation}

\subsection{Covariance between $\gamma_t(r_p)$ and $w_{gg}(r_p)$}

For the cross-probe covariance between the galaxy-galaxy lensing and the galaxy-galaxy projected correlation function, we have
\begin{align}
    \text{Cov}&(\gamma_t(r_i), w_{gg}(r_j)) \nonumber\\
	= &\int \int \frac{ dk_1^3 }{ (2\pi)^3 } \, 
    \frac{ dk_2^3 }{ (2\pi)^3 }\, e^{i\bm{k_1}\cdot\bm{r_i}} \, e^{i\bm{k_2}\cdot\bm{r_j}} \, \, \text{Cov}^{2D}\left(P_{g \gamma}(k_1), \, P_{gg}(k_2)\right) \nonumber\\
    = \, &\frac{2}{V_s} \left[ \int d\Pi \, W_{g \gamma}(\Pi) \, W_{gg}(\Pi) \right] \nonumber\\
	&\times \int_0^{\infty} \frac{k \, dk}{2\pi} \, J_2(k r_i) J_0(k r_j) \left[ 
            P_{gg}(k)
        \,  P_{\gamma g}(k) \right] \nonumber\\
    \approx \, &\frac{4 \Pi_{\text{max}}}{V_s} \int_0^{\infty} \frac{k \, dk}{2\pi} \, J_2(k r_i) J_0(k r_j) \nonumber\\
     &\times \left( \frac{\bar\rho}{\Sigma_c(\chi_{l}, \chi_{s})} \right) \, \Pgm(k) \, \Pgg(k) \, .
\end{align}

\subsection{Projected radius of a survey}

For various computations related to the covariance matrix (e.g., the effective source density $\Sigma_s$), it is necessary to compute the effective projected radius. For this purpose, we assume a survey at a single redshift $z$ with spherical cap geometry in a flat universe, where the effective survey radius $R_s$ is such that the area of a flat circle with radius $R_s$ is equivalent to that of a spherical cap survey. Then we have
\begin{align}
\pi R_s^2 = 4 \pi \chi^2 f_\text{sky} \, ,
\end{align}
where $f_\text{sky}$ is the fraction of the sky covered by the survey and $\chi$ is the comoving radial distance to the effective survey redshift $z$. Then we have
\begin{align}
R_s = 2 \chi \sqrt{f_{\text{sky}}} \, .
\end{align}
For a survey of 9736 sq. deg. at an effective redshift $z = 0.3$, $R_s \approx 812.9 \, \hMpc$.

\section{Gaussian process implementation}
\label{appendix:GP}

The Gaussian process predictor for the expected value of scalar-valued process (i.e., function) $y(\v{x_{\star}})$ is
\begin{align}
\hat y(\v{x}_{\star}) = \sum_{i}^{N} k(\v{x}_i, \v{x}_{\star}) \, \alpha_i \, ,
\end{align}
where the coefficients $\alpha_i$ are computed by
\begin{align}
\v{\alpha} = (K + \sigma_{ii}^2 I)^{-1} \v{y} \, ,
\label{eq:gp_inverse}
\end{align}
where $K_{ij} = k(\v{x}_i, \v{x}_j)$, $\v{y}$ is the vector of training data observations, $k$ is the kernel (i.e., covariance) function, and the indices $i$,$j$ run over the training data of size $N$. This is simply a Wiener filter applied to the training data \citep{Rybicki_1992}. As such, a key assumption of this method is that noise of the process that we seek to predict is Gaussian.  In our application, the training data consists of the values of $w_{\text{p,sim}}/w_{\text{p,analytic}}$ (or $\Delta\Sigma$) in a given radial bin and the parameter values at which these were computed. We estimate the noise $\sigma_{ii}^2$ of each datapoint via the sample variance estimator of the 20 stochastic HOD realizations computed at each value of the parameters.

The surge in recent applications is due to the realization that such a smoothing filter can be applied to arbitrary machine learning (i.e., curve fitting) problems by marginalizing over the unknown function $k$ \citep{Ohagan_1978,Sacks_1989,Rasmussen_2006}.  Due to computational expense, and as is standard practice, one can find a single covariance function $k$ (over some well-defined class of functions) which maximizes the marginal likelihood of the Gaussian process.

However, instead of maximizing the marginal likelihood itself, we instead maximize a closely-related function, namely, the leave-one-out \emph{pseudo}-likelihood
\begin{align}
&\ln \mathcal{L} = \sum_{i}^{N} \left[ -\frac{1}{2} \log \left( \lambda \sigma_i^2 \right) - \frac{1}{2}
  \frac{||\text{LOOE}||_i^2}{\lambda \sigma_i^2} - \frac{1}{2} \log 2\pi \right] \, , \\
\label{eq:GP_likelihood}
&||\text{LOOE}||_i^2 = \left( \hat y_{(i), i} - y_{i} \right)^2 \, ,
\end{align}
where $N$ is the number of training points, $||\text{LOOE}_i||$ is the
leave-one-out cross-validation error on training point $i$, $\sigma_i^2$ is the noise variance of
 on training point $i$, and $\lambda$ is a free parameter that rescales all of the variances. This
function is obtained by taking the product over $i=1,2,\dots,N$ of the conditional likelihoods for
training point $y_i$ given that all training points \emph{except} $y_i$ are known, which is precisely
the situation when performing leave-one-out cross-validation.  This \emph{pseudo}-likelihood has the
useful property that it is independent of the form of the kernel function $k$, unlike
the true marginal likelihood of the Gaussian process (which depends on the log determinant of $K$),
which ensures that the maximum of this function attempts to minimize the prediction error regardless of the complexity of the kernel function \citep{Rasmussen_2006}.

In order to efficiently compute the leave-one-out predictions $\hat y_{(i)}$, we use the identity for the inverse of a $2 \times 2$ block matrix (e.g., \citealt{Press_1992}), assuming without loss of generality that $i = N$. In this case, this identity reduces to the covariance matrix $K$ partitioned into sub-blocks, with $A$ an $(N-1) \times (N-1)$ matrix, $b$ an $(N-1)$ column vector, and $c$ a scalar:
\begin{align}
K^{-1} = 
\begin{bmatrix}
A 	& b \\
b^T & c
\end{bmatrix}^{-1} = 
\begin{bmatrix}
\tilde A	& \tilde b \\
\tilde b^T	& \tilde c
\end{bmatrix} \, ,
\end{align}
where
\begin{align}
\tilde A &= (A - c^{-1} b b^T)^{-1} = A^{-1} + \tilde c A^{-1} b b^T A^{-1} \\
\tilde b &= -\tilde c A^{-1} b \\
\tilde c &= (c - b^T A^{-1} b)^{-1} \, .
\end{align}
With repeated application of the Sherman-Morrison-Woodbury formula, we then solve for $A^{-1}$ in terms of $\tilde b$, $b$ and $\tilde A$ to obtain
\begin{align}
A^{-1} = \left( I - \tilde b b^T \right) \, \tilde A \, .
\end{align}
Recognizing that $\hat y_{ (i) } = b^T A^{-1} \v{y}_{ (i) }$, we obtain the leave-one-out prediction
\begin{align}
\hat y(x_i)_{ (i) } &= - b^T b \tilde b^T \tilde A \v{y}_{ (i) } \, ,
\end{align}
which can be rewritten in terms of the true value $y_i$, the matrix elements of $K^{-1}$, and the matrix elements of $K^{-1} \v{y}$ as (\citealt{Rasmussen_2006}, Eq. 5.12)
\begin{align}
\hat y(x_i)_{ (i) } &=  y_i - \frac{[K^{-1} \v{y}]_i}{[K^{-1}]_{ii}} \, .
\end{align}
The leave-one-out error on point $i$ is therefore
\begin{align}
||\text{LOOE}||_i =  \frac{[K^{-1} \v{y}]_i}{[K^{-1}]_{ii}} \, .
\end{align}

For our kernel function $k$, we adopt the squared exponential family of kernel functions
\begin{align}
k(\v{x}_i,\v{x}_j) = \sigma^2_{\text{signal}} \exp{ \left( -\frac{1}{2} \, \v{d}^{T} \Lambda \v{d} \right) } + \sigma^2_{\text{mean}} \, ,
\end{align}
where $\v{d} = \v{x}_i - \v{x}_j$, $\Lambda$ is a diagonal matrix of hyperparameters over which to optimize, 
and $\sigma^2_{\text{signal}}$ and $\sigma^2_{\text{mean}}$ are hyperparameters over which to optimize.
This kernel is a convenient choice of an infinitely differentiable and translation invariant function
and is perhaps the most common kernel used in machine learning applications \citep{Rasmussen_2006}.

To optimize the hyperparameters ($\Lambda$, $\sigma^2_{\text{signal}}$, $\sigma^2_{\text{mean}}$, $\lambda$),  we use
the \textsc{BOBYQA} derivative-free quadratic surface optimization method \citep{Powell_2009} included in the
\textsc{nlopt} software package \citep{nlopt}. We find that the performance of the emulator is very sensitive
to the initial guess of the hyperparameters, and that an initial guess for the hyperparameters
which implies a high signal-to-noise of the training data is crucial in order to avoid being
trapped in a sub-optimal local maximum of eq. \ref{eq:GP_likelihood}. In practice, we accomplish this by de-dimensionalizing all inputs and outputs of the emulator, rescaling the range of each input parameter $x_i$ to $[0, 1]$ and rescaling the training data outputs $y_i$ so that $y_i$ has mean zero and variance unity. Then we choose an initial guess for $\lambda \equiv 0.01$ and $\sigma_{\text{signal}} \equiv 1$, which implies that the training data have a signal-to-noise $S/N$ greater than implied by the input uncertainties by a factor $\lambda^{-1/2} = 10$. For our training data, even allowing $\lambda$ to vary, this is sufficent to ensure that the optimization finds a relatively high signal-to-noise solution rather than running away toward a low signal-to-noise solution with low predictive power.

\section{Projection integrals with finite bin size}
\label{appendix:projection_integrals}

In order to take advantage of statistical isotropy to increase the signal-to-noise of our simulation training data, we directly measure the real-space (cross-)correlation functions, and then transform to projected quantities. A slight complication arises when computing these from simulations rather than from theory since we must tabulate these quantities as bin-averaged correlation functions when counting pairs of particles in simulations, but the projection integrals are defined in terms of the raw correlation functions.

We compute the projection integrals from the tabulated correlation functions as follows. Starting from the Abel integral transform
\begin{align}
w_{\text{p,xy}}(r_p) = 2 \int_{r_p}^{\Pi_{\text{max}}} \xi_{\text{xy}} \left( \sqrt{r_p^2 + \Pi^2} \right) \, d\Pi \, ,
\end{align}
we assume $\xi_{\text{xy}}$ is tabulated such that correlation function values are exact at midpoints of the tabulated bins $r_i$ (a more accurate approach would be to compute the mean pair-weighted separation within each bin). We then analytically integrate, using piecewise linear elements on the interval between the midpoints of adjacent tabulated bins [$r_{i,-}$, $r_{i,+}$], to obtain a second-order accurate (in adjacent bin separation $\Delta r$) sum for $w_p(r_p)$:
\begin{align}
w_{p}(r_p) &= \sum_i 2 \, \left( \xi_{i,-} - m_i r_{i,-} \right) \left( \sqrt{ s_{i,+}^2 - r_p^2 } - \sqrt{ s_{i,-}^2 - r_p^2 } \right) \nonumber\\
&+ \sum_i m_i \left[ s_{i,+} \sqrt{ s_{i,+}^2 - r_p^2 } \, - \,  s_{i,-} \sqrt{ s_{i,-}^2 - r_p^2 } \right. \nonumber\\
& \left. + \, r_p^2 \ln \left( \frac{ s_{i,+} + \sqrt{ s_{i,+}^2 - r_p^2 } }{ s_{i,-} + \sqrt{ s_{i,-}^2 - r_p^2 } } \right) \, \right] \, ,
\end{align}
where
\begin{align}
m_i &= \frac{\xi_{i,+} - \xi_{i,-}}{r_{i,+} - r_{i,-}}
\end{align}
and
\begin{align}
s_{i,-} &= \max \left( r_p, \, 	 r_{i,-} \right) \\
s_{i,+} &= \min \left( \Pi_{\text{max}}, \, r_{i,+} \right) \, .
\end{align}


\bsp	
\label{lastpage}
\end{document}

%% file: header.tex
\renewcommand{\v}[1]{\ensuremath{\mathbf{#1}}} 
 
\let\baraccent=\= 
\renewcommand{\=}[1]{\stackrel{#1}{=}} 


%% file: Posteriors/lowz_allruns_posterior_table.tex
\begin{tabular}{l@{\quad}ll@{\quad}ll@{\quad}ll@{\quad}ll@{\quad}ll@{\quad}ll@{\quad}ll}
\toprule
{Parameter} & {Fiducial} & {Lower $f_{\text{sat}}$} & {Higher $f_{\text{sat}}$} & {15\% Incompl.} & {`Baryons'} & \textbf{LOWZ} & $\bm{[> 2 \, h^{-1} \, \text{\textbf{Mpc}}]}$ \\
\midrule
$n_{g} \times 10^{4}$ & $ 2.90^{+0.19}_{-0.29}$ & $ 3.00\pm 0.23$ & $ 3.13^{+0.32}_{-0.14}$ & $ 3.04^{+0.34}_{-0.22}$ & $ 2.98\pm 0.23$ & $ 3.08^{+0.38}_{-0.16}$ & $ 2.97^{+0.25}_{-0.31}$ &  \\ 
$M_0 / M_1$ & $ < 0.174$ & $ 0.20^{+0.11}_{-0.13}$ & $ < 0.161$ & $ 0.178^{+0.073}_{-0.16}$ & $ < 0.161$ & $ < 0.131$ & $ 0.20\pm 0.10$ &  \\ 
$M_1 / M_{\text{min}}$ & $ 12.8^{+2.0}_{-3.9}$ & $ 11.9^{+1.3}_{-2.8}$ & $ 12.0^{+1.6}_{-3.2}$ & $ 11.08^{+0.73}_{-3.3}$ & $ 13.0^{+2.1}_{-3.9}$ & $ < 11.2$ & $ 13.5^{+2.9}_{-3.8}$ &  \\ 
$\sigma_{\log M}$ & $ 0.165^{+0.049}_{-0.15}$ & $ < 0.211$ & $ 0.174^{+0.053}_{-0.16}$ & $ 0.207^{+0.099}_{-0.14}$ & $ < 0.207$ & $ 0.291^{+0.11}_{-0.082}$ & $ 0.186^{+0.081}_{-0.13}$ &  \\ 
$\alpha$ & $ 0.90^{+0.19}_{-0.27}$ & $ < 0.921$ & $ 0.82^{+0.14}_{-0.25}$ & $ 0.85^{+0.13}_{-0.30}$ & $ 0.89^{+0.17}_{-0.27}$ & $ < 0.831$ & $ 1.04^{+0.34}_{-0.24}$ &  \\ 
$A_{\text{conc}}$ & $ > 1.47$ & $ 1.74\pm 0.65$ & $ 1.71^{+0.69}_{-0.83}$ & $ 1.79^{+0.82}_{-0.68}$ & $ ---$ & $ 1.78^{+0.82}_{-0.69}$ & $ 1.70^{+0.65}_{-0.78}$ &  \\ 
$R_{\text{rescale}}$ & $ 0.91\pm 0.14$ & $ 0.82^{+0.13}_{-0.12}$ & $ 1.03\pm 0.16$ & $ 0.87^{+0.17}_{-0.15}$ & $ 0.92^{+0.16}_{-0.13}$ & $ 1.17^{+0.17}_{-0.095}$ & $ > 1.18$ &  \\ 
\midrule
$S_8$ & $ 1.019^{+0.040}_{-0.055}$ & $ 1.017\pm 0.032$ & $ 0.985^{+0.036}_{-0.043}$ & $ 0.981\pm 0.032$ & $ 1.006^{+0.038}_{-0.049}$ & $ 0.847\pm 0.037$ & $ 0.853^{+0.034}_{-0.057}$ &  \\ 
$\sigma_8$ & $ 0.818^{+0.036}_{-0.054}$ & $ 0.826\pm 0.033$ & $ 0.803^{+0.036}_{-0.045}$ & $ 0.798^{+0.030}_{-0.035}$ & $ 0.808^{+0.036}_{-0.052}$ & $ 0.692^{+0.014}_{-0.037}$ & $ < 0.707$ &  \\ 
$\Omega_m$ & $ 0.324^{+0.023}_{-0.0089}$ & $ 0.318^{+0.020}_{-0.016}$ & $ 0.316\pm 0.014$ & $ 0.317^{+0.020}_{-0.016}$ & $ > 0.317$ & $ 0.315^{+0.020}_{-0.017}$ & $ 0.313^{+0.019}_{-0.016}$ &  \\ 
$H_0$ & $ 67.1^{+2.2}_{-3.1}$ & $ 67.5^{+2.4}_{-2.7}$ & $ 70.3^{+4.4}_{-0.96}$ & $ 68.1\pm 2.7$ & $ 66.9^{+2.1}_{-3.4}$ & $ 67.7^{+3.3}_{-4.3}$ & $ 67.1^{+2.3}_{-4.4}$ &  \\ 
$n_s$ & $ 0.961^{+0.021}_{-0.016}$ & $ 0.960\pm 0.016$ & $ 0.959^{+0.016}_{-0.019}$ & $ 0.957^{+0.013}_{-0.021}$ & $ 0.9559^{+0.0096}_{-0.024}$ & $ > 0.961$ & $ 0.962^{+0.021}_{-0.013}$ &  \\ 
$\Omega_b$ & $ 0.0505^{+0.0049}_{-0.0039}$ & $ 0.0493\pm 0.0042$ & $ > 0.0535$ & $ 0.0490\pm 0.0044$ & $ 0.0498^{+0.0051}_{-0.0042}$ & $ 0.0485^{+0.0043}_{-0.0064}$ & $ 0.0473^{+0.0023}_{-0.0076}$ &  \\ 
$w_0$ & $ -1.02^{+0.21}_{-0.18}$ & $ -0.98^{+0.22}_{-0.16}$ & $ -0.97^{+0.24}_{-0.15}$ & $ -1.00^{+0.21}_{-0.18}$ & $ -1.03\pm 0.18$ & $ -1.02\pm 0.16$ & $ -1.01^{+0.19}_{-0.17}$ &  \\ 
$A_{\text{lensing}}$ & $ 1.020\pm 0.049$ & $\text{N/A}$ & $\text{N/A}$ & $\text{N/A}$ & $ 1.039\pm 0.048$ & $ 0.970^{+0.048}_{-0.043}$ & $ 0.967\pm 0.048$ &  \\ 
\bottomrule
\end{tabular}